# Sensitive bioassay with an ultra-large dynamic range via microlaser ensemble quenching

Weishu Wu[1,2,3], Yuhang Cao[1,2,3], Xiaotian Tan[4], and Xudong Fan[1,2,3,*]

[1]Department of Biomedical Engineering,
University of Michigan, Ann Arbor, MI 48109, USA

[2]Center for Wireless Integrated MicroSensing and Systems (WIMS[2]),
University of Michigan, Ann Arbor, MI 48109, USA

[3]Max Harry Weil Institute for Critical Care Research and Innovation,
University of Michigan, Ann Arbor, MI 48109, USA

[4]Shenzhen Institute of Advanced Technology,
Chinese Academy of Science, Shenzhen, Guangdong 518071, P. R. China

*Corresponding author: xsfan@umich.edu

**Abstract**

We present a bioassay platform that leverages the lasing threshold distribution in a microlaser ensemble (ME), consisting of hundreds of individual microlasers, to measure analyte concentrations in solution. An ME is formed by placing dye-doped microbeads in a micro Fabry-Perot cavity. Microbeads are surface modified with biorecognition molecules to capture analytes, while the quenchers resulting from the presence of the analytes on the microbeads' surfaces increase the lasing thresholds of microlasers. Since the number of analytes varies from one microbead (or microlaser) to another due to the randomness in binding processes, a distribution of the analytes (and hence the quenchers) in the ME is created, which in turn leads to a lasing threshold distribution in the ME. Experimentally, multiple pumping energy densities are used to probe the lasing threshold distribution. A theoretical model is developed to map the lasing threshold distribution to analyte distribution in the ME, and then to recover the analyte concentration in solution. Using streptavidin and interleukin-6 as a model system, our platform achieves a detection limit of 0.1 pg/mL and a dynamic range exceeding five orders of magnitude, showing that the ME quenching method can provide a high sensitivity with a superior dynamic range.

**Keywords**

**Introduction**

Bioassays are to quantify target analytes (such as biomolecules) in a sample. Their core working principle is to convert the presence of the target analytes into measurable signals. Among the various methods, enzyme-linked immunosorbent assay (ELISA) is a commonly used technique[1,2], in which detection antibodies conjugated with enzymes bind to the captured target analytes and the enzyme-substrate reaction causes the substrate to emit light (such as chemiluminescence)[3,4] or change its color (colorimetric detection)[5,6]. Alternatively, detection antibodies can be conjugated with fluorescent labels[7] such as dyes[8], quantum dots[9-11], and plasmonic nanoparticles (*i.e.*, metal nanoparticles)[12-14]. The light signal from these labels is measured to quantify the target analyte concentration. Since the measured signal varies continuously with the concentration of the analyte, these traditional approaches are referred to as analog detection. Such methods are highly refined, widely available in commercial kits, and typically achieve detection ranges from a few picograms per milliliter (pg/mL) to tens of thousands of pg/mL, depending on the specific assay formats, reagents, and target analytes[15-19].

To further improve the sensitivity (or detection limit) of bioassays, digital ELISA was developed[20-27]. This approach enables unprecedented sensitivity by allocating individual analyte molecules to an ensemble of microunits (such as microbeads) and then counting the fraction of these microunits that emit light (bright microunits). Under the assumption that each microunit has no more than one analyte molecule associated with it, detection of the analyte concentration becomes analyzing the distribution of the bright microunits and calculating its mean value according to the Poisson distribution. Commercial examples, such as Quanterix's bead-based digital ELISA, can detect cytokines as low as 0.01 pg/mL[20,28-30]. Other works have employed alternative labeling methods, such as plasmonic nanoparticles[31,32], fluorescent dyes[33,34], and quantum dots[35,36], to replace the enzyme labels[23,25-27]. While these digital immunoassays can achieve unprecedented sensitivity, they have a limited dynamic range. For example, the upper detection limit of Quanterix's bead-based method is only 24 pg/mL[29,30]. At higher concentrations, the assumption for digital ELISA, that is, the average number of analytes per microbead should be far below 1, breaks down[37]. In order to extend the dynamic range of digital ELISA, samples under test need to be serially diluted so that the concentration of the target analytes falls in the range allowed by digital ELISA. However, the appropriate dilution factor needs to be determined through multiple trials. In addition, for multiplexed detection involving multiple analytes with vastly different concentrations, it may be difficult to find the one-size-fits-all dilution factor that works for all analytes. Other strategies have also been explored, including stitching the digital detection calibration curve with the analog detection calibration curve[21,27] or extrapolating the calibration curve beyond the single molecule assumption[38]. However, these attempts face issues. It is difficult to determine the cutoff (or transition) analyte concentration between the digital detection mode and analog detection mode[21,27] and there may be discontinuity in digital and analog calibration curves due to two completely different methods used to obtain digital and analog sensing signal[38]. Furthermore, extrapolating the calibration curve leads to large errors. Therefore, a unified method that has a high dynamic range to cover both the lower and the upper end of the analyte concentration without any artificially introduced digital-to-analog transition is highly desirable.

In this work, we propose a microlaser ensemble (ME) quenching bioassay platform that achieves both a high sensitivity and a large dynamic range. In this method, individual microlasers are used as detection microunits. First, these microlasers are exposed to the analytes in solution and subsequently capture the analytes through biochemical interactions (such as antibody-antigen binding). The presence of the analytes in these microlasers alters their lasing thresholds through the quenchers produced by enzyme-substrate reactions. When the external pump varies, the laser emission from some of the microlasers in the ME persists, whereas other microlasers are completely quenched, depending on the number of quenchers (and hence the analytes) associated with each microlaser. As a result, the distribution of the analytes on the ME can be established, which allows us to recover the analyte concentration in solution through a statistical model. This platform achieves a detection limit of 0.1 pg/mL and a dynamic range spanning more than five orders of magnitude without any calibration curve stitching/extrapolation or sample dilution. We foresee that this highly sensitive platform with a large dynamic range and the related statistical method will find broad real-world applications

**Working Principle**

Without losing generality, in this work, we use a microlaser formed by placing a dye-doped microbead along with a liquid medium into a micro Fabry-Perot (F-P) cavity (Fig. 1(a)). Each microbead, together with the F-P cavity, is treated as an individual detection microunit. Other forms of the microunits such as ring resonator lasers[39,40], nanowire lasers[41], and vertical-cavity surface-emitting lasers (VCSELs) can also be used.

A single microlaser formed by a dye-doped microbead in an F-P cavity has been extensively studied previously by our group[42]. When a microlaser is pumped by an external laser at an energy density higher than its lasing threshold, it emits laser signals. We refer to it as a bright microlaser, as illustrated in Fig. 1(a). For the same microlaser, when it captures analytes, which results in light absorbing molecules (quenchers) to be near or on the microbead surface (through a process such as enzyme-substrate reaction), the lasing threshold increases. When the lasing threshold surpasses the external pumping energy density, the microlaser is quenched (a dark microlaser), as illustrated in Fig. 1(b).

We now move from a single microlaser to an ME, which is composed of many identical microlasers. In a quencher-free state, when an ME is pumped at an energy density slightly higher than a microlaser's lasing threshold, all microlasers are bright (Fig. 1(c)). When the ME goes through analyte-capturing and quencher-generating processes, the randomness of the analyte binding to the microbeads leads to a distribution of the analytes on the microlasers, which in turn causes a distribution of quenchers and hence lasing thresholds within the ME. Under a given external pumping energy density, only a portion of the microlasers, whose lasing thresholds are lower than the pumping energy density, remain bright, as shown in Fig. 1(d).

Fig. 1(e) presents schematically the distribution of the lasing threshold. For the case of Fig. 1(c) (quencher-free), the lasing threshold for all microlasers is the same and low, as represented by the δ-function like green line on the left. Correspondingly, the number of quenchers on the microlaser is zero and its distribution is a δ function (the green line in Fig. 1(f)). For the case of Fig. 1(d), the initial δ-function like threshold distribution in Fig. 1(e) starts to increase and spread. The microlasers whose lasing thresholds are on the left side of the distribution curve (*i.e.*, the lasing thresholds are lower than the pumping energy density) remain bright, whereas those on the right side (*i.e.*, the lasing thresholds are higher than the pumping energy density) become dark (or quenched). By changing the external pumping energy density and counting the fraction of the bright microlasers (which is the cumulative distribution of the lasing threshold), we can map the distribution of the lasing threshold and hence the underlying distribution of quenchers on the microlasers shown in Fig. 1(f). With the known quencher distribution, we can recover the analyte distribution on the microlasers and hence the analyte concentration in solution through a statistical model.

**Theoretical Model**

We developed the microlaser quenching equation (MQE) for the ME to study the relationship between the lasing thresholds of microlasers and the number of analytes they capture. The MQE, as shown in Eq. (1), is built on well-recognized theoretical models for microlasers[43-46] and an assumption that the quencher production rate is proportional to the number of enzymes, which is in turn proportional to the number of analytes captured by the microlasers. The equation:

$$k_j = \frac{ln\left(\frac{I_{th(j)}^{exp} + I_{norm}}{I_0 + I_{norm}}\right)}{E} \qquad (1)$$

states that for the $j_{th}$ microlaser in the ensemble with the quencher-free lasing threshold, $I_0$, its experimentally measured lasing threshold, $I_{th(j)}^{exp}$, can be used to estimate the number of analytes $k_j$. $E$, $I_0$, and $I_{norm}$ are the three fitting parameters determined by the experimental setup. The detailed derivation of Eq. (1) can be found in Section S1 of the Supplementary Information.

When the ME is exposed to the analytes in solution, the analyte molecules bind to the microlasers through interactions such as antigen-antibody binding. The binding processes are affected by the random movement of the analyte molecules, leading to a distribution of the analytes on the microlasers, which can be experimentally measured in the following way.

When the ME is pumped at multiple pumping energy densities, a data pair $[(Lf(i), I_{pump}^{exp}(i))]$ can be obtained, where $Lf(i)$ is the lasing fraction of the ME at the pumping energy density of $I_{pump}^{exp}(i)$. As discussed in the Working Principle section, the lasing fraction is equivalent to the fraction of microlasers with the number of analytes smaller than a specific value.

In our experiments, the number of microlasers is always lower than the number of analytes in the test sample, that is, the average number of analytes per microlaser is larger than one. From the central limit theorem, we infer that the distribution of the analytes on the microlasers follows a Gaussian distribution. Therefore, from Eq. (1), the lasing fraction of the ME, $Lf(i)$, and its paired measured pumping energy density, $I_{pump}^{exp}(i)$, should follow a Gaussian cumulative distribution:

$$Lf(i) = \phi_{cumu}\left(\frac{ln\left(\frac{I_{pump}^{exp}(i) + I_{norm}}{I_0 + I_{norm}}\right)}{E}; \mu, \sigma\right), \qquad (2)$$

where $\phi_{cumu}$ is the cumulative Gaussian distribution function. $\mu$ and $\sigma$ are the mean and standard deviation for this Gaussian distribution, respectively, which can be obtained via function fitting. $\mu$

is related to the analyte concentration in solution. Such a relationship can be established by calibration curves obtained experimentally, which will be discussed later in the Results section.

Probing the distribution of the analyte on microlasers (or, more essentially, on microbeads inside the micro F-P cavity) can also be accomplished with simple fluorescence intensity measurements rather than using the proposed microlaser method (or lasing threshold analysis). However, as theoretically analyzed in Section S2, the lasing threshold analysis method is much more sensitive and accurate in obtaining the analyte distribution among the microbeads.

**Materials and Methods**

(1) ME quenching bioassay protocol

We present a bioassay protocol for the ME quenching experiments in Fig. 2. As shown in the top panel of Fig. 2, before the microbeads were placed into an F-P cavity to form the ME, they underwent a surface modification step, an analyte capture step, and a labeling step. First, dye-doped microbeads (mean diameter: 5 μm, FCDG008, Bangs Laboratories Inc., USA) surface modified and conjugated with the desired capture antibodies using a conjugation kit (PL01N, Bangs Laboratories Inc., USA) and following manufacturer's protocols. The microbeads were then incubated with SuperBlock$^{TM}$ (37515, Thermo Fisher, USA) for 60 minutes followed by two repeated wash (PBS) steps. The microbeads were gently rotated during each incubation step using a rotator (R2020, Benchmark Scientific, USA). For the analyte capture step, the test sample was mixed with functionalized microbeads and incubated for 60 minutes followed by two repeated wash (PBS) steps. For the labeling step, the microbeads were incubated with biotinylated detection antibodies for 60 minutes, followed by two repeated wash (PBS) steps, and then with 1:250 diluted streptavidin poly-HRP (21140, Thermo Fisher, USA) for 30 min followed by three repeated wash steps (PBS). The microbead suspension was centrifuged at 400 g for 15 minutes to complete liquid exchange. For IL-6 ELISA experiments, antibody sets from a test kit were used (DY206, R&D Systems, USA).

Finally, these microbeads, now with captured analytes and labeled with poly-HRP, were mixed with precipitative substrate solution and immediately loaded into an F-P cavity to form the ME (see the lower panel of Fig. 2). The F-P cavity consisted of two highly reflective mirrors (Evaporated Coatings, Inc., USA)[47,48], each of 1" by 1". The distance of the F-P cavity was controlled by the dye-doped microbeads (*i.e.*, 5 μm). We used a 1:1 mixture of 10x concentrated metal-enhanced DAB (34065, Thermo Fisher, USA) and concentrated hydrogen peroxide (34062, Thermo Fisher, USA) as the substrate in this procedure. The mixture was incubated in the F-P cavity for 30 minutes so that abundant precipitates that act as quenchers for microlasers could form. Although for each analyte concentration, ~10000 microbeads were used to incubate with the sample and subsequently loaded into the F-P cavity. In the measurement and data analysis step, we processed only ~1000 microlasers located at the center of the F-P cavity. Therefore, the number of microlasers in an ME was ~1000.

We also used a simpler biotin-streptavidin binding assay to evaluate our platform. In this simplified assay, biotinylated antibodies (MQ2-39C3, Thermo Fisher, USA) were conjugated to the surface of the microbeads using the same conjugation kit mentioned previously (PL01N, Bangs Laboratories Inc., USA). Streptavidin conjugated with poly-HRP served as the analyte, which was captured by biotin on microbeads through 20 minutes of incubation. The microbeads were then washed three times and mixed with the substrate solution in the same manner as the standard ME quenching immunoassay protocol described above before they were loaded into an F-P cavity.

(2) Lasing threshold measurements of the ME

In the measurement step, the ME (*i.e*., an F-P cavity with microbeads and quenchers in it) was placed under a custom-built laser emission microscope where an external nanosecond laser source (473 nm, 5 kHz repletion rate) was used to scan over the microlasers, and the images were taken. Details of the laser emission microscope can be found in our previous publications[47,48]. A white light image of the microbeads within the F-P cavity was taken to exactly count the total number of microlasers. The ME was pumped at six pumping energy densities (25 µJ/mm$^2$, 45 µJ/mm$^2$, 90 µJ/mm$^2$, 125 µJ/mm$^2$, 250 µJ/mm$^2$, and 400 µJ/mm$^2$) and the corresponding lasing fraction of the ME for each pumping energy density was recorded.

**Results**

(1) Demonstration of microlaser quenching

We first investigated the quenching of microlasers by enzyme-substrate reactions and optimized the assay protocol to determine the enzyme-substrate incubation time. A biotin-streptavidin binding assay using a fixed 1 ng/mL of streptavidin poly-HRP was performed. The four sets of MEs were prepared in the same manner according to the assay procedures described in the Materials and Methods section, except that incubation time with substrate was chosen to be 0 minutes, 10 minutes, 20 minutes, and 30 minutes, respectively. Each set of the ME was scanned at six selected pumping energy densities.

Fig. 3(a) shows an exemplary microlaser from each set of the ME with a fixed incubation time, illustrating how the increased incubation time affects the microlasers. As incubation time increases from 0 minutes (bottom row) to 30 minutes (top row), the microlasers' lasing thresholds increase progressively, meaning that they may be quenched more easily at a lower pumping energy density. Fig. 3(b) provides more quantitative analysis of the four sets of the MEs. It shows that the MEs with longer incubation times exhibit lower lasing fractions, as more quenchers can form. In subsequent experiments, we chose an incubation time of 30 minutes to achieve a strong quenching effect. Note that the incubation time here refers to the enzyme-substrate reaction time, which is an additional period on top of sample loading and other procedure-related time during the experiment.

From Fig. 3(b), we can see that the lasing fraction is lower than 100%, even at a high pumping energy density (400 $\mu J/mm^2$), which suggests that there exist "bad" microlasers whose gains cannot overcome the cavity loss and/or the loss caused by quenchers. While the exact reasons for "bad" microlasers are unknown, we speculate that this could be due to some microbeads having a very low dye doping density or whose shape deviates far from spherical that causes a higher cavity loss. The effect of these "bad" microlasers (*i.e.*, those that do not lase at the highest pumper energy density, 400 $\mu J/mm^2$ in this work) will be removed in our data analysis in the next section by normalization.

(2) A biotin-streptavidin binding assay

We tested the ME quenching bioassay using a biotin-streptavidin binding assay. The protocol is a simplified version of immunoassays described in the Materials and Methods section. After conjugation of biotinylated antibodies, the microbeads were separated into 13 equal portions, including 12 positive samples, incubated in 12 five-fold, serial-diluted streptavidin poly-HRP solutions and 1 negative control sample (containing 0 pg/mL streptavidin poly-HRP). For each of the 13 portions of microbeads, we created a corresponding set of ME by placing the microbeads in an F-P cavity, as described in the Materials and Methods.

Figure 4(a) shows the lasing fraction of the 13 sets of MEs under various pumping energy densities. It is seen that for a given set of ME that corresponds to a fixed analyte concentration, the lasing fraction increases with the increased pumping energy density. If we regroup these datapoints based on each pumping energy density, which are plotted in six colored curves, it is seen that within each curve (*i.e.*, under the same pumping energy density) the lasing fraction decreases with the increased analyte concentration and hence increased number of quenchers associated with the microlasers. As discussed in Fig. 3, there exist "bad" microlasers that may not lase even at the highest pumping energy density (400 µJ/mm$^2$ in our case) that may affect our lasing fraction calculation. To remove the impact of the "bad" microlasers, in Fig. 4(b) we normalize each lasing fraction curve to the corresponding lasing fraction obtained with the negative control, that is, when the analyte (streptavidin) concentration is 0 pg/mL (or 10$^{-5}$ pg/mL in the logarithmic scale). Details of the normalization description can be found in the Section S3. From here on, we will use the normalized lasing fraction data.

From Fig. 4(b), we can see that when those microlasers are incubated with analytes of various concentrations, the (normalized) lasing fraction curves that originally converge at 0 pg/mL start to diverge. For example, at 0.1 pg/mL, the lasing fraction is about 73% for the pumping energy density of 25 µJ/mm$^2$, whereas the lasing fraction remains nearly 100% for the pumping energy density of 400 µJ/mm$^2$, which indicates that the number of quenchers (hence the number of analytes) varies from one microlaser to another, which results in a lasing threshold distribution among the microlasers. When different pumping energy densities are used for a given analyte concentration, we probe the analyte distribution on the ME, which has previously been discussed in the Working Principle section. Note that in a hypothetical scenario in which the number of quenchers (or the number of analytes) is the same for all microlasers, the lasing fraction curves should collapse to a single curve (*i.e.*, all curves are completely overlapped).

The lasing fraction curves in Fig. 4(b) can be treated as a group of calibration curves analogous to the calibration curve used in the conventional ELISA. The differences are that in our method, we have multiple calibration curves (six in Fig. 4(b)), whereas in the conventional ELISA, only one calibration curve is available. Note that in conventional fluorescence based immunoassays, multiple calibration curves can be generated via different pumping power as well. However, due to the linear nature of the fluorescence intensity with respect to the pumping power, all calibration curves provide the same information and can be reduced to a single calibration curve after linear rescaling (see the illustration in Fig. S3). In contrast, due to the nonlinear nature of the laser that has threshold behavior, the calibration curves in Fig. 4(b) are linearly independent and fundamentally determined by the analyte distribution on the microlasers. These multiple calibration curves work together to help pinpoint the analyte concentration more precisely and enable a larger dynamic range than a single curve in the conventional ELISA.

(3) Analyte concentration recovery using the MQE

The MQE presented in Eq. (1) and the Gaussian distribution of the quenchers (analytes) on microlasers establish the connection between the lasing fraction and three variables: the pumping energy density and the two Gaussian parameters (the mean value µ and the standard deviation σ) of the distribution of the analytes on the ME, which is reflected in Eq. (2). Furthermore, the average number of analytes in the ME, µ, is positively related to the analyte concentration in solution ($C$), which is similar to digital ELISA where the average number of bright microunits is related to the analyte concentration in solution. Since µ is an intermediate parameter, eventually, the relationship among the lasing fraction, pumping energy density, and the analyte concentration $C$, which parameters are all experimentally measurable, can be established by calibration curves obtained experimentally using function fitting of lasing fraction of the ME with respect to the experimental pumping energy density using Eq. (2).

In the fitting process, the three constants in Eq. (2), $E$, $I_0$, and $I_{norm}$, are held constant for each set of ME data, since these constants are related to the experimental conditions such as enzyme/substrate type and microlaser properties, which are the same for all sets of MEs. For each set of ME incubated in one analyte concentration $C_h$, two Gaussian parameters, $\mu_h$ and $\sigma_h$ can be used to describe its lasing fraction distribution. In the data fitting process, the Gaussian parameter pair for all sets of ME used in fitting ($\mu_h$ and $\sigma_h$, where $h$ runs from 1 to $M$ – the number of total sets of the MEs used in data fitting) and the three aforementioned constants ($E$, $I_0$, and $I_{norm}$) are fitted together to minimize the overall loss for all sets of the MEs. This fitting process guarantees that each data set shares the same parameters that describes the system, namely, $E$, $I_0$, and $I_{norm}$.

In our work, the function fitting (lasing fraction vs. pumping energy density) is performed for the nine sets of data (*i.e.*, $M = 9$) in Fig. 4(b) with the analyte concentration ranging from $4\times10^4$ pg/mL to $1\times10^{-1}$ pg/mL, as shown in the subfigures of Fig. 5(a) along with the corresponding fitted values of µ and σ (*i.e.*, $\mu_{fit}$ and $\sigma_{fit}$). During the above function fitting, we drop three sets of data associated with excessively high analyte concentrations ($5\times10^6$ pg/mL, $1\times10^6$, and $2\times10^5$ pg/mL), since their corresponding laser fractions are extremely low at all pumping energy densities and provide too few data points for a reliable Gaussian fit. For example, only one microlaser is bright in the $5 \times 10^6$ pg/mL set of ME. Note that $\mu_{fit}$ is not necessarily the actual average number of quenchers (or analytes) of the ME, since it is subject to a rescaling factor, *i.e.*, the fitting constant $E$. This is because $E$ is an unbounded parameter and contains information of quencher/enzyme (analyte) ratio during the incubation step. The exact value of $E$ is determined by a microscopic quencher producing equation and it is unknown to us. However, from Eq. (2) we can see that while $E$ is unbounded, the product, $E\mu$, is always fixed. Therefore, depending on the value of $E$ that we choose, $\mu_{fit}$ varies and is rescaled by $E$. In our current work, $E$ is set to be one ($E = 1$).

To quantitatively recover the analyte concentration, we introduce a new parameter, the recovered lasing threshold, $I_\mu$, which is connected to $\mu$ via the following equation.

$$\mu = \frac{ln\left(\frac{I_\mu + I_{norm}}{I_0 + I_{norm}}\right)}{E}, \qquad (3)$$

or

$$I_\mu = (I_0 + I_{norm})e^{\mu E} - I_{norm} . \qquad (4)$$

Note that the invariable product of $E\mu$ is used in Eq. (4) and therefore the $I_\mu$ value is unique. $I_\mu$ can be deemed as the lasing threshold that corresponds to an imaginary microlaser with the averaged number ($\mu$) of quenchers. Fig. 5(b) plots the calibration curve for $I_\mu$ vs. the analyte concentration, showing good linearity in the logarithmic scale.

We can use the above calibration curve to obtain the analyte concentration for a sample under test, we will first measure the lasing fraction of an ME at different pumping energy densities. Then lasing fraction vs. pumping energy density will be fitted to the Gaussian cumulative distribution function, Eq. (2), where the parameter, $\mu$, can be calculated (the values of $E$, $I_0$, and $I_{norm}$ will remain the same as those used during calibration curve generation). Then the calculated $\mu$ will be transformed to an $I_\mu$ value using Eq. (4). Finally, the analyte concentration can be recovered by checking the calculated $I_\mu$ value with the calibration curve in Fig. 5(b). $I_\mu$ is obtained through the fitting of an ME over multiple pumping energy densities (six in our current work) for significantly improved stability and robustness of measurements over single-point measurement (such as fluorescence measurement).

Based on the linearity of $I_\mu$ and analyte concentration in the logarithmic scale, we then performed 3D fitting of the lasing fraction vs. the analyte concentration vs. the pumping energy density. The fitted surface is presented in Fig. 5(c), showing the continuous change of the lasing fraction with respect to analyte concentration and pumping energy density. Figs. 5(b) and (c) show that our system is continuously responsive over concentration change of more than five orders of magnitude and achieves a low detection limit of 0.1 pg/mL.

Note that this 3D surface also works as a calibration surface (in 3D) that functions like a calibration curve in conventional ELISA (2D). The 3D calibration surface allows for pinpointing the concentration of the analyte under test using only one pumping energy density. In this case, the sample under test will incubate with an ME, then the lasing fraction will be measured at only one pumping energy density. A point on the calibration surface will be located with the measured lasing fraction under the given pumping energy density, which provides the value of the analyte concentration. Although a single-point measurement is faster and simpler than the function-fitting method that we demonstrated here, it lacks stability and may lead to large measurement errors.

(4) IL6 immunoassay

We further demonstrated highly sensitive detection of interleukin-6 (IL-6) with the large dynamic range using our ME quenching bioassay platform. The assay protocol is described in the Materials and Methods section. In IL-6 assay, we had seven sets of MEs, including six positive sets incubated in 6 ten-fold, serial-diluted IL-6 solutions (from 0.01 pg/mL to 1000 pg/mL) and one negative control sample (containing 0 pg/mL IL-6).

First, we present the function fitting of lasing fraction with respect to pumping energy density for each set of ME in Fig. 6(a). The fitting process is the same as the one used in Fig. 5(a). Three constants $E$, $I_0$, and $I_{norm}$ are held constant for all six sets of MEs. Using Eq. (5), we obtain the calibration curve of the recovered lasing threshold ($I_\mu$) with respect to analyte concentration ($C$), as shown in Fig. 6(b). Our method achieves a detection limit of 0.1 pg/mL for IL-6 detection and remains sensitive at 1 ng/mL, achieving a dynamic range of four orders of magnitude.

Figure 6(c) shows the 3D plot of datapoints of lasing fraction change with respect to analyte concentration and pumping energy density. The fitted surface is obtained using the linear relationship between $I_\mu$ and analyte concentration in the logarithmic scale in Fig. 6(b). This surface shows that our system is continuously responsive over four orders of magnitude. Similar to Fig. 5(c), this surface plot can also function as a single-point measurement calibration surface.

## Discussion

In this work, we proposed and demonstrated a new bioassay platform that uses microlaser ensembles to probe the distribution of analytes among the microlasers and subsequently deduce the analyte concentration in solution using a statistical model. Our method is fundamentally different from digital ELISA.

The limitation in digital detection is that while it can distinguish one analyte from zero analytes for each microunit, it is unable to differentiate one from two or more analytes in each microunit. As illustrated in Fig. S4(a), the presence of two or more analytes in a microunit saturates the system. Therefore, digital ELISA relies on a fundamental assumption that the average number of analytes per microunit is far below one. When the average number of analytes per microunit approaches one for the entire ensemble, significant deviations arise. Consequently, digital ELISA has a limited dynamic range (especially in the upper range of detection).

In contrast, our microlasers, as microunits, have tunable dynamic ranges, which can be achieved by tuning external pumping. This concept is mathematically described in Eq. (2) and schematically illustrated in Fig. S4(b), which allows our system to achieve a large dynamic range. The current assumption in our method is that the number of analytes exceeds the number of microunits (microlasers in our current work), so that we can use the Gaussian distribution. In the future, when even lower analyte concentration is pursued, a new statistical model can be developed that may provide an even more sensitive detection and larger dynamic range.

Furthermore, both digital ELISA and conventional ELISA provide only one calibration curve. In digital ELISA, it is the average number of analytes vs. analyte concentration. In conventional ELISA, it is detection signal (such as light intensity or absorption vs. analyte concentration) vs. analyte concentration. In contrast, due to the threshold behavior intrinsic to lasers, our method has multiple calibration curves. By fitting multiple nonlinearly related data points to the multiple calibration curves, our method significantly improves measurement stability and robustness in concentration recovery when compared to both digital and conventional ELISA.

Essentially, our method is to map the analyte distribution among detection microunits (microlasers in our case). This is similar to digital ELISA, which also maps the distribution of the analytes in an ensemble of microunits but at a very low average number of analytes per microunits. Alternately, we can use the conventional fluorescence based method to detect the light intensity from individual detection microunit to map the distribution, which, unlike digital ELISA, does not require the average number of analytes per microunit be far below one. However, according to our theoretical analysis in Section S2 in the Supplementary Information, our microlaser quenching method has a sensitivity approximately six orders of magnitude greater than that of a fluorescence based method, Therefore, our method is more accurate in obtaining the analyte distribution information.

While a single microlaser has previously been used as a biosensor to recover the analyte concentration through laser intensity measurement[49] or laser onset time measurement[50], these single-point measurements do not provide stable and robust data, as discussed previously. Moreover, the laser onset time method[50] requires extremely long assay time and the long time exposure to external pumping light may bleach the gain medium, leading to reduced sensitivity and a large variation.

**Conclusion**

In this work, we presented the theory and experimental realization of the microlaser ensemble quenching bioassay platform that is fundamentally different from digital ELISA. Using streptavidin and IL-6 as a model system, we demonstrated that this system is sensitive, achieving a detection limit of 0.1 pg/mL and can potentially have a dynamic range of exceeding five orders of magnitude. Our system uses the assumption of Gaussian distribution for analyte numbers and thus can effectively recover high analyte concentrations. Our platform has an advantage in diagnostic applications where test-sample biomarkers can span multiple orders of magnitude.

In the future, we plan to extend the dynamic range in both the low and high end of analyte concentration. To detect lower concentrations of analytes, microbeads doped with a lower density of dyes can be used. Alternatively, quenchers with a higher quenching strength can be used. Both approaches will make the microlaser more sensitive to the quenching effect caused by a lower number of analytes (or quenchers) associated with the microlaser. More homogeneous microlasers can also be explored to reduce the lasing fraction variances in measurements, thus helping the system to reach lower concentrations. To detect higher concentrations of analytes, larger pumping energy densities ($>400$ µJ/mm$^2$) can be used to pump the ME, so that we can observe more bright microlasers to perform function fitting. Another research direction that can be pursued is spectral-multiplexed detection that takes advantage of narrow lasing emission spectral linewidth, which allows the detection of different lasing wavelengths from different dyes within even a single-color channel (*e.g*., green color)[51]. Finally, some other types of microlasers can be explored, such as ring resonators, which do not rely on external F-P cavities, and VCSELs, which are made of semiconductor materials with high uniformity (*i.e*., microlaser homogeneity) and can be pumped electrically.

**Acknowledgement**

The authors thank the support from Richard A. Auhll Professorship from the University of Michigan.

**Conflict of Interest**

X.F. is an inventor of the laser emission technologies used in this work, which are licensed to LEMX Health Technology Co., LTD. He and the University of Michigan have financial interests in LEMX.

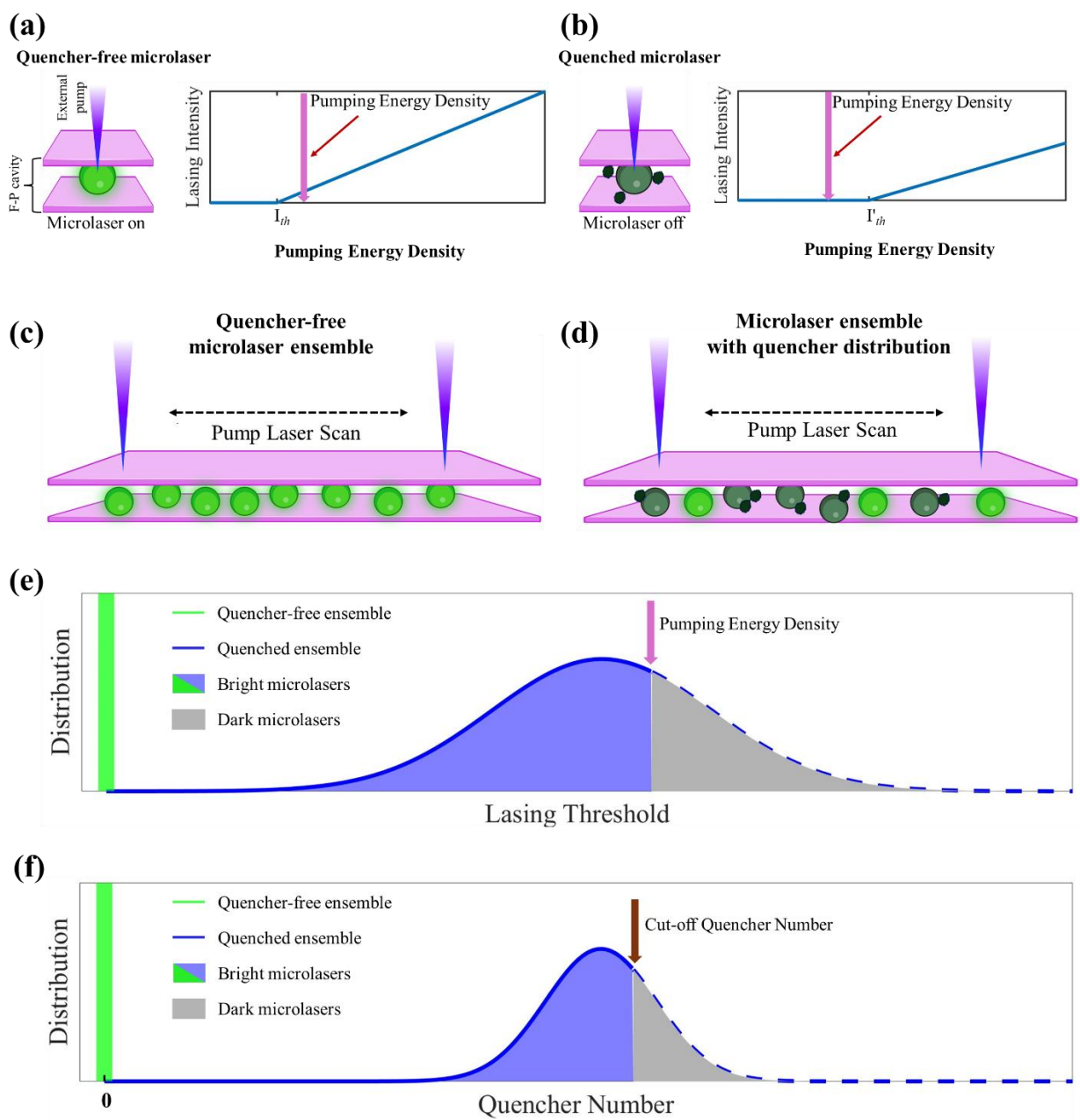

**Figure 1.** Conceptual illustration of the microlaser ensemble (ME) quenching bioassay. (a) A single microlaser is formed by placing a microbead doped with laser gain medium (such as dye) into a liquid-filled micro Fabry–Perot (F-P) cavity (liquid medium not shown for clarity). The microlaser emits laser signal (bright microlaser) when the external pump exceeds its lasing threshold. (b) If the microbead has quenchers near or on its surface due to the analytes captured by the microbead, its lasing threshold can surpass the pumping energy density, thus turning off the laser emission (dark microlaser). (c) A microlaser ensemble refers to a group of identical dye-

doped microbeads deposited in the same F-P cavity. When the external pump exceeds the lasing threshold, all the microlasers in the ME emit laser signals (bright microlasers). (d) When the ME is exposed to the sample under test, due to the randomness of the analyte binding to the microlasers, different microlasers have different numbers of analytes (and hence quenchers) associated with them. Consequently, the lasing threshold of each microlaser increases according to the number of quenchers near or on its surface, thus creating a lasing threshold distribution within the ME. Under the same external pump, some microlasers are quenched (dark microlasers) whereas the rest of microlasers still have laser emission (bright microlasers). (e) Illustration of measuring the lasing threshold distribution in an ME. Quencher-free microlasers have the same lasing thresholds as shown by the d-function-like green line on the left. The presence of quencher distribution in the microlaser ensemble leads to a distribution of the lasing threshold as represented by the blue curve. Under a given pumping energy density, the microlasers in the blue-shaded area, which have the lasing threshold lower than the pumping energy density, are bright microlasers, whereas the microlasers in the gray-shaded area are quenched microlasers (dark microlasers). By counting the bright microlasers within the ensemble when the pumping energy density is varied, the lasing threshold distribution can be mapped. (f) Mapping the lasing threshold distribution is equivalent to mapping the quencher distribution and hence the analyte distribution within the microlaser ensemble, which is in turn related to the analyte concentration in solution. The d-function-like green line on the left represents the situation where none of the microlasers in the ensemble has a quencher on or near the microbead surface.

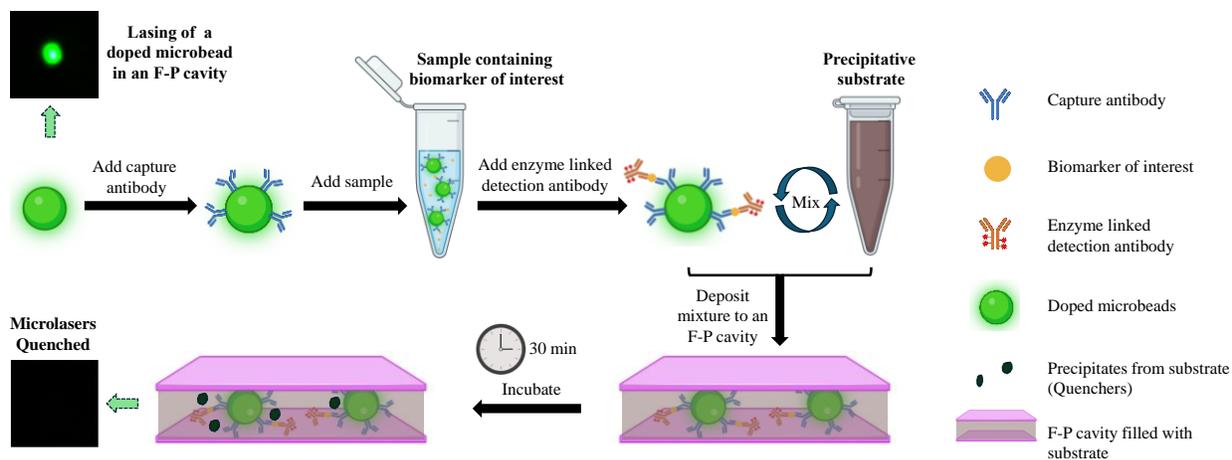

**Figure 2.** Illustration of the microlaser ensemble quenching bioassay. For clarity, only one or two microlasers are used to represent the process for the whole ensemble. First, dye-doped microbeads conjugated with capture antibodies are incubated with the sample under test. After the biomarkers are captured, enzyme-linked detection antibodies are added and bound to the microbeads, which are subsequently mixed with substrate solution. The mixture is deposited into an F-P cavity, where enzyme-substrate reaction produces precipitations that work as quenchers for the microlasers, producing a distribution of the lasing threshold. This lasing threshold distribution can be measured by changing the external pumping energy density and subsequently used to calculate the analyte concentration. The image on the top left corner shows laser emission from a microlaser (*i.e.*, dye-doped microbead in an F-P cavity). The image on the bottom left corner shows the laser emission from the microlaser is quenched.

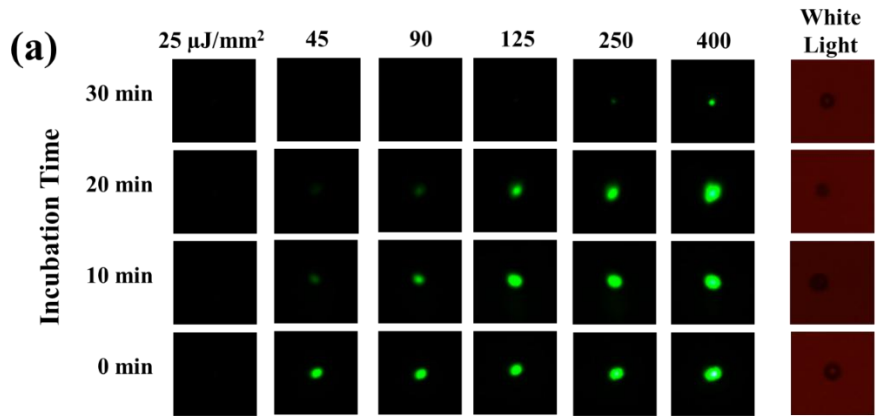

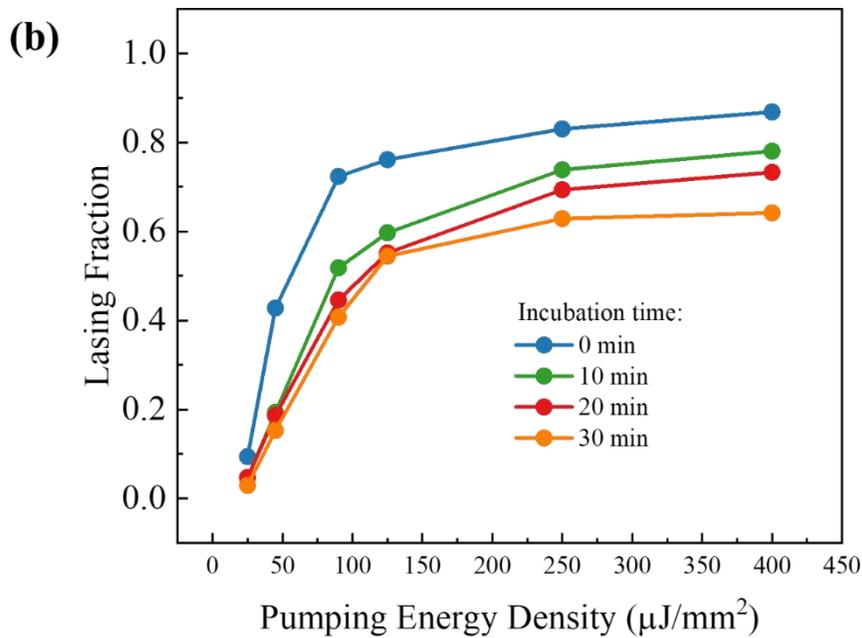

**Figure 3.** (a) Example of a microlaser at various pumping energy densities over various incubation times. Each image has dimensions of 35 μm × 35 μm. The rightmost column contains white light images of the microlaser (microbead). (b) Change in the lasing fraction within the same microlaser ensemble with respect to incubation time. Quenchers formed within the microlasers over time gradually increase microlasers' lasing thresholds, thus gradually reducing the lasing fraction in the ME under a fixed pumping energy density. Streptavidin poly-HRP = 1 ng/mL.

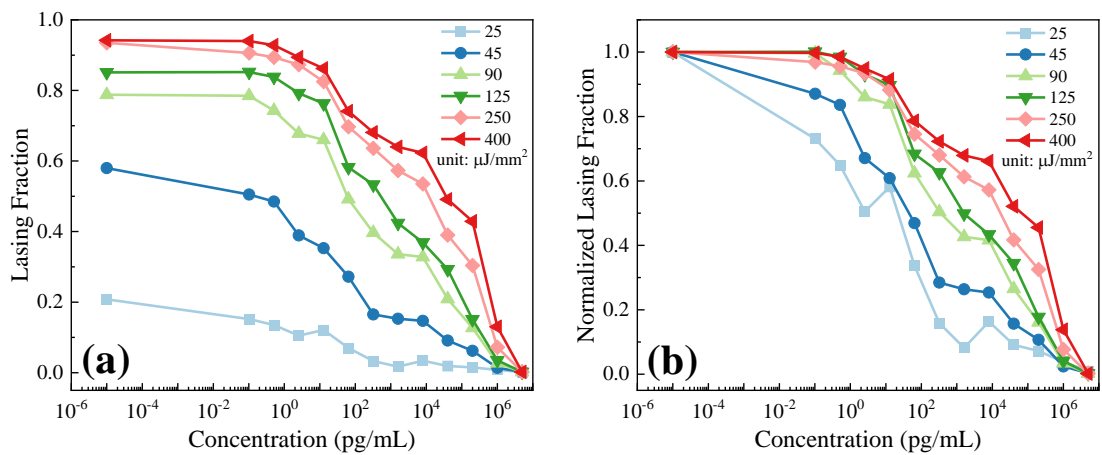

**Figure 4.** Lasing fraction of the MEs with respect to the analyte (streptavidin) concentration under various pumping energy densities. (a) Raw data. (b) Normalized to zero analyte concentration for each pumping energy density. Note that in the logarithmic scale in the x-axis, we use $10^{-5}$ pg/mL to represent zero analyte concentration.

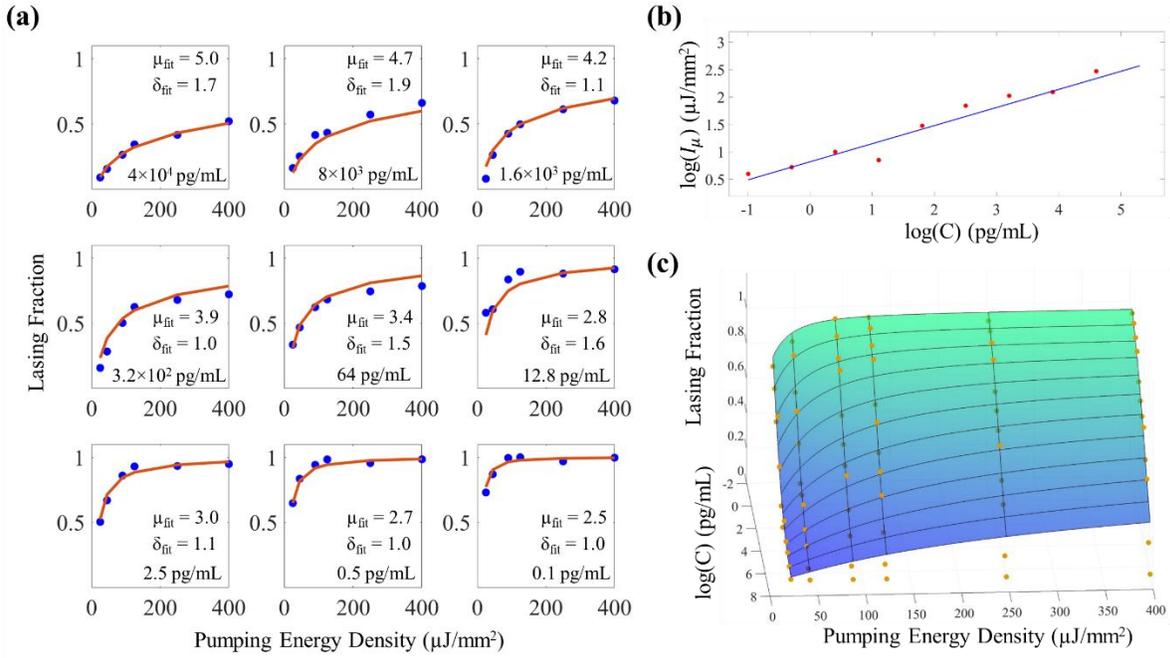

**Figure 5.** (a) Calibration curves obtained using function fitting of lasing fraction vs. pumping energy density for the nine sets of data in Fig. 4. Each subfigure represents the lasing fraction of an ME with respect to pumping energy density for a given analyte concentration in solution ($C$). The function fitting is performed using Eq. (2). $\mu_{fit}$ and $\sigma_{fit}$ refer to the fitted values for the Gaussian distribution. For all nine sets of data, $I_{norm}$ and $I_0$ are calculated to be 3.4 μJ/mm² and 1.1 μJ/mm², respectively. $E$ is set to 1 for the fitting. Note that the above function fitting, the three sets of data from the three highest concentrations (5×10⁶ pg/mL, 1×10⁶, and 2×10⁵ pg/mL) are not included, since their lasing fractions are close to zero, which do not provide enough information for the Gaussian distribution. (b) Calibration lines for $I_\mu$ vs. analyte concentration in the log-log scale. (c) 3D fitted surface plot of the lasing fraction of the microlaser ensemble with respect to pumping energy density and analyte (streptavidin) concentration. The surface fitting is shown with scattered experimental data (dots). Here, log($C$) refers to the logarithmic analyte concentration in units of pg/mL. Note that some datapoints appear darker than others, meaning that they are below the surface, whereas others are on or above the surface.

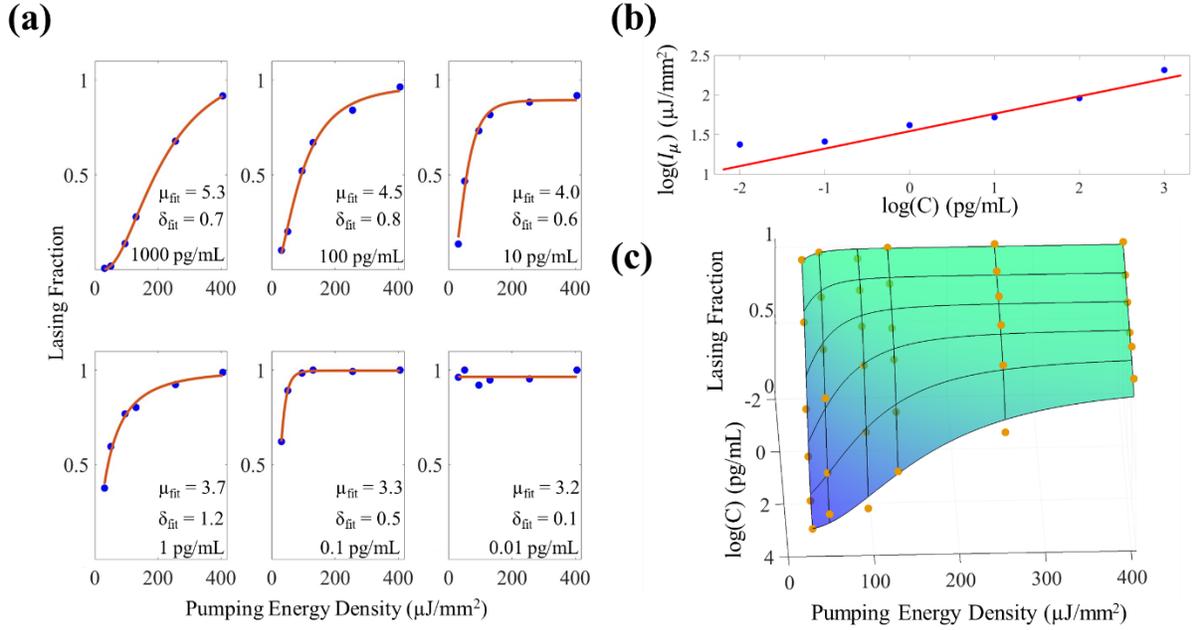

**Figure 6.** (a) Calibration curves obtained using function fitting of lasing fraction vs. pumping energy density for the six sets of experimental data. Each subfigure represents the lasing fraction of an ME with respect to pumping energy density for a given analyte concentration in solution ($C$). The function fitting uses Eq. (2). $\mu_{fit}$ and $\sigma_{fit}$ refer to the fitted values for the Gaussian distribution. For all six sets of data, $I_{norm}$ and $I_0$ are calculated to be 2.1 μJ/mm$^2$, and $I_0$ is calculated to be 0.1 μJ/mm$^2$, respectively. $E$ is set to 1 for the fitting. (b) Calibration lines for $I_\mu$ vs. analyte concentration in the log-log scale. (c) 3D fitted surface plot of the lasing fraction of the microlaser ensemble with respect to pumping energy density and analyte (IL-6) concentration. The surface fitting is shown with scattered experimental data (dots). Here, log($C$) refers to the logarithmic analyte concentration in units of pg/mL. Note that some datapoints appear darker than others, meaning that they are below the surface, whereas others are on or above the surface.

## References

1. D. Wild, *The immunoassay handbook: theory and applications of ligand binding, ELISA and related techniques* (Newnes, 2013).

2. D. S. Hage, "Immunoassays," Anal. Chem. **71**, 294-304 (1999).

3. I. Surugiu, B. Danielsson, L. Ye, K. Mosbach, and K. Haupt, "Chemiluminescence imaging ELISA using an imprinted polymer as the recognition element instead of an antibody," Anal. Chem. **73**, 487-491 (2001).

4. M. Herrmann, T. Veres, and M. Tabrizian, "Enzymatically-generated fluorescent detection in micro-channels with internal magnetic mixing for the development of parallel microfluidic ELISA," Lab Chip **6**, 555-560 (2006).

5. K. Pei, Y. Xiong, B. Xu, K. Wu, X. Li, H. Jiang, and Y. Xiong, "Colorimetric ELISA for ochratoxin A detection based on the urease-induced metallization of gold nanoflowers," Sens. Actuators, B **262**, 102-109 (2018).

6. S. Chen, M. Svedendahl, R. P. Van Duyne, and M. Käll, "Plasmon-enhanced colorimetric ELISA with single molecule sensitivity," Nano Lett. **11**, 1826-1830 (2011).

7. C. Hempen and U. Karst, "Labeling strategies for bioassays," Anal. Bioanal. Chem. **384**, 572-583 (2006).

8. C. M. O'Donnell and S. C. Suffin, "Fluorescence immunoassays," Anal. Chem. **51**, 33A-40A (1979).

9. E. R. Goldman, I. L. Medintz, and H. Mattoussi, "Luminescent quantum dots in immunoassays," Anal. Bioanal. Chem. **384**, 560-563 (2006).

10. R. Thürer, T. Vigassy, M. Hirayama, J. Wang, E. Bakker, and E. Pretsch, "Potentiometric immunoassay with quantum dot labels," Anal. Chem. **79**, 5107-5110 (2007).

11. Y. Lv, F. Wang, N. Li, R. Wu, J. Li, H. Shen, L. S. Li, and F. Guo, "Development of dual quantum dots-based fluorescence-linked immunosorbent assay for simultaneous detection on inflammation biomarkers," Sens. Actuators, B **301**, 127118 (2019).

12. M. Seydack, "Nanoparticle labels in immunosensing using optical detection methods," Biosens. Bioelectron. **20**, 2454-2469 (2005).

13. N. Kongsuwan, X. Xiong, P. Bai, J.-B. You, C. E. Png, L. Wu, and O. Hess, "Quantum plasmonic immunoassay sensing," Nano Lett. **19**, 5853-5861 (2019).

14. J. Ling, Y. F. Li, and C. Z. Huang, "Visual sandwich immunoassay system on the basis of plasmon resonance scattering signals of silver nanoparticles," Anal. Chem. **81**, 1707-1714 (2009).

15. M. Helle, L. Boeije, E. de Groot, A. de Vos, and L. Aarden, "Sensitive ELISA for interleukin-6: detection of IL-6 in biological fluids: synovial fluids and sera," J. Immunol. Methods **138**, 47-56 (1991).

16. J. Gratacos, A. Collado, X. Filella, R. Sanmarti, J. Canete, J. Llena, R. Molina, A. Ballesta, and J. Muñoz-Gómez, "Serum cytokines (IL-6, TNF-α, IL-1β and IFN-γ) in ankylosing spondylitis: a close correlation between serum IL-6 and disease activity and severity," Rheumatology **33**, 927-931 (1994).

17. S. A. Khan, J. Joyce, and T. Tsuda, "Quantification of active and total transforming growth factor-β levels in serum and solid organ tissues by bioassay," BMC Res. Notes **5**, 1-9 (2012).

18. S. Jayasena, M. Smits, D. Fiechter, A. De Jong, J. Nordlee, J. Baumert, S. L. Taylor, R. H. Pieters, and S. J. Koppelman, "Comparison of six commercial ELISA kits for their specificity and sensitivity in detecting different major peanut allergens," J. Agric. Food Chem. **63**, 1849-1855 (2015).

19. A. A. Al-Hosary, J. Ahmed, A. Nordengrahn, and M. Merza, "Assessment of the first commercial ELISA kit for the diagnosis of theileria annulata," J. Parasitol. Res. **2015**, 787812 (2015).

20. D. M. Rissin, C. W. Kan, T. G. Campbell, S. C. Howes, D. R. Fournier, L. Song, T. Piech, P. P. Patel, L. Chang, A. J. Rivnak, E. P. Ferrell, J. D. Randall, G. K. Provuncher, D. R. Walt, and D. C. Duffy, "Single-molecule enzyme-linked immunosorbent assay detects serum proteins at subfemtomolar concentrations," Nat. Biotechnol. **28**, 595-599 (2010).

21. J. Hwang, M. Banerjee, A. S. Venable, Z. Walden, J. Jolly, C. Zimmerman, E. Adkisson, and Q. Xiao, "Quantitation of low abundant soluble biomarkers using high sensitivity single molecule counting technology," Methods **158**, 69-76 (2019).

22. S. A. Byrnes, T. Huynh, T. C. Chang, C. E. Anderson, J. J. McDermott, C. I. Oncina, B. H. Weigl, and K. P. Nichols, "Wash-free, digital immunoassay in polydisperse droplets," Anal. Chem. **92**, 3535−3543 (2020).

23. L. Cohen, N. Cui, Y. Cai, P. M. Garden, X. Li, D. A. Weitz, and D. R. Walt, "Single molecule protein detection with attomolar sensitivity using droplet digital enzyme-linked immunosorbent assay," ACS Nano **14**, 9491-9501 (2020).

24. Y. Song, Y. Ye, S.-H. Su, A. Stephens, T. Cai, M.-T. Chung, M. K. Han, M. W. Newstead, L. Yessayan, D. Frame, H. D. Humes, B. H. Singer, and K. Kurabayashi, "A digital protein microarray for COVID-19 cytokine storm monitoring," Lab Chip **21**, 331-343 (2021).

25. Z. Gao, Y. Song, T. Y. Hsiao, J. He, C. Wang, J. Shen, A. MacLachlan, S. Dai, B. H. Singer, K. Kurabayashi, and P. Chen, "Machine-learning-assisted microfluidic nanoplasmonic digital immunoassay for cytokine storm profiling in COVID-19 patients," ACS Nano **15**, 18023−18036 (2021).

26. C. Wu, T. J. Dougan, and D. R. Walt, "High-throughput, high-multiplex digital protein detection with attomolar sensitivity," ACS Nano **16**, 1025-1035 (2022).

27. C. Chen, R. Porter, X. Zhou, C. L. Snozek, E. H. Yang, and S. Wang, "Microfluidic digital immunoassay for point-of-care detection of NTproBNP from whole blood," Anal. Chem. **96**, 10569−10576 (2024).

28. https://www.quanterix.com/.

29. https://www.quanterix.com/product-brochures/ss-simoa-il-17a-advantage-plus/.

30. https://www.quanterix.com/product-brochures/ss-measure-ultra-low-levels-of-ifn-%CE%B1-with-simoa-advantage-plus/.

31. A. Belushkin, F. Yesilkoy, and H. Altug, "Nanoparticle-enhanced plasmonic biosensor for digital biomarker detection in a microarray," ACS Nano **12**, 4453-4461 (2018).

32. W. Zhang, T. Dang, Y. Li, J. Liang, H. Xu, G. L. Liu, and W. Hu, "Digital plasmonic immunosorbent assay for dynamic imaging detection of protein binding," Sens. Actuators, B **348**, 130711 (2021).

33. S.-M. Yang, Q. Bi, W. J. Zhang, X. Cui, Y. Zhou, C. Yuan, and Y. Cui, "Highly accurate multiprotein detection on a digital ELISA platform," Lab Chip **22**, 3015-3024 (2022).

34. D. M. Rissin, C. W. Kan, L. Song, A. J. Rivnak, M. W. Fishburn, Q. Shao, T. Piech, E. P. Ferrell, R. E. Meyer, and T. G. Campbell, "Multiplexed single molecule immunoassays," Lab Chip **13**, 2902-2911 (2013).

35. X. Liu, Y. Sun, X. Lin, X. Pan, Z. Wu, and H. Gai, "Digital duplex homogeneous immunoassay by counting immunocomplex labeled with quantum dots," Anal. Chem. **93**, 3089-3095 (2021).

36. Q. Zhang, J. Li, X. Pan, X. Liu, and H. Gai, "Low-numerical aperture microscope objective boosted by liquid-immersed dielectric microspheres for quantum dot-based digital immunoassays," Anal. Chem. **93**, 12848-12853 (2021).

37. Y. Zhang and H. Noji, "Digital bioassays: theory, applications, and perspectives," Anal. Chem. **89**, 92-101 (2017).

38. J. Zhang, A. D. Wiener, R. E. Meyer, C. W. Kan, D. M. Rissin, B. Kolluru, C. George, C. I. Tobos, D. Shan, and D. C. Duffy, "Improving the accuracy, robustness, and dynamic range of digital bead assays," Anal. Chem. **95**, 8613-8620 (2023).

39. Q. Chen, H. Liu, W. Lee, Y. Sun, D. Zhu, H. Pei, C. Fan, and X. Fan, "Self-assembled DNA tetrahedral optofluidic lasers with precise and tunable gain control," Lab Chip **13**, 3351–3354 (2013).

40. N. Martino, S. J. J. Kwok, A. C. Liapis, H.-M. Kim, S. J. Wu, J. Wu, S. Forward, H. Jang, P. H. Dannenberg, S.-J. Jang, Y.-H. Lee, and S.-H. Yun, "Wavelength-encoded laser particles for massively multiplexed cell tagging," Nat. Photonics **13**, 720-727 (2019).

41. X. Wu, Q. Chen, P. Xu, Y.-C. Chen, B. Wu, R. M. Coleman, L. Tong, and X. Fan, "Nanowire lasers as intracellular probes," Nanoscale **10**, 9729-9735 (2018).

42. X. Wu, Y. Wang, Q. Chen, Y.-C. Chen, X. Li, L. Tong, and X. Fan, "High-Q, low-mode-volume microsphere-integrated Fabry-Pérot cavity for optofluidic lasing applications," Photon. Res. **7**, 50-60 (2019).

43. M. Aas, Q. Chen, A. Jonáš, A. Kiraz, and X. Fan, "Optofluidic FRET lasers and their applications in novel photonic devices and biochemical sensing," IEEE J. Sel. Top. Quantum Electron. **22**, 188-202 (2015).

44. S. Lacey, I. M. White, Y. Sun, S. I. Shopova, J. M. Cupps, P. Zhang, and X. Fan, "Versatile opto-fluidic ring resonator lasers with ultra-low threshold," Opt. Express **15**, 15523-15530 (2007).

45. H.-J. Moon, Y.-T. Chough, and K. An, "Cylindrical microcavity laser based on the evanescent-wave-coupled gain," Phys. Rev. Lett. **85**, 3161 (2000).

46. A. E. Siegman, *Lasers* (University Science Books, 1986).

47. Q. Chen, Y.-C. Chen, Z. Zhang, B. Wu, R. Coleman, and X. Fan, "An integrated microwell array platform for cell lasing analysis," Lab Chip **17**, 2814-2820 (2017).

48. W. Wu, Y. Zhang, X. Tan, Y. Chen, Y. Cao, V. Sahai, N. Peterson, L. Goo, S. Fry, and V. Kathawate, "Antigen-independent single-cell circulating tumor cell detection using deep-learning-assisted biolasers," Biosens. Bioelectron., 116984 (2024).

49. C. Gong, Y. Gong, M. K. K. Oo, Y. Wu, Y. Rao, X. Tan, and X. Fan, "Sensitive sulfide ion detection by optofluidic catalytic laser using horseradish peroxidase (HRP) enzyme," Biosens. Bioelectron. **96**, 351-357 (2017).

50. X. Wu, M. K. K. Oo, K. Reddy, Q. Chen, Y. Sun, and X. Fan, "Optofluidic laser for dual-mode sensitive biomolecular detection with a large dynamic range," Nat. Commun. **5**, 3779 (2014).

51. Y.-C. Chen, X. Tan, Q. Sun, Q. Chen, W. Wang, and X. Fan, "Laser-emission imaging of nuclear biomarkers for high-contrast cancer screening and immunodiagnosis," Nat. Biomed. Eng. **1**, 724-735 (2017).

# Sensitive bioassay with an ultra-large dynamic range via microlaser ensemble quenching

Weishu Wu[1,2,3], Yuhang Cao[1,2,3], Xiaotian Tan[4], and Xudong Fan[1,2,3],*

[1]Department of Biomedical Engineering,

University of Michigan, Ann Arbor, MI 48109, USA

[2]Center for Wireless Integrated MicroSensing and Systems (WIMS[2]),

University of Michigan, Ann Arbor, MI 48109, USA

[3]Max Harry Weil Institute for Critical Care Research and Innovation,

University of Michigan, Ann Arbor, MI 48109, USA

[4]Shenzhen Institute of Advanced Technology,

Chinese Academy of Science, Shenzhen, Guangdong 518071, P. R. China

*Corresponding author: xsfan@umich.edu

# S1. Theoretical Model

## S1.1    Single microlaser case

In order for a microlaser to lase, the following condition should be met[1-4].

$$n_{1,Gain}\sigma_{e,Gain} \geq \left(n_{T,Gain}-n_{1,Gain}\right)\sigma_{a,Gain} + L + n_q\sigma_{a,q}, \qquad (1)$$

where $n_{T,Gain}$ is the total density of the laser gain molecules in the microlaser. $n_{1,Gain}$ is the density of the laser gain molecules in the excited state. $\sigma_{e,Gain}$ and $\sigma_{a,Gain}$ are the emission and absorption cross section of the laser gain molecules at the lasing wavelength, respectively. L is the intrinsic per single-trip laser cavity loss in the absence of quenchers. $n_q$ and $\sigma_{a,q}$ are the quencher density and quencher's absorption cross section at the lasing wavelength. Re-arranging Eq. (1), we have

$$\frac{n_{1,Gain}}{n_{T,Gain}} \geq \frac{\sigma_{a,Gain}}{\sigma_{e,Gain}+\sigma_{a,Gain}}[1+\frac{L}{n_{T,Gain}\sigma_{a,Gain}}] + \frac{n_q\sigma_{a,q}}{n_{T,Gain}(\sigma_{e,Gain}+\sigma_{a,Gain})}, \qquad (2)$$

which can be approximated as

$$\frac{n_{1,Gain}}{n_{T,Gain}} \geq \frac{\sigma_{a,Gain}}{\sigma_{e,Gain}}[1+\frac{L}{n_{T,Gain}\sigma_{a,Gain}}] + \frac{n_q\sigma_{a,q}}{n_{T,Gain}\sigma_{e,Gain}}, \qquad (3)$$

because $\sigma_{a,Gain}$ ($\sim 10^{-18}$ cm$^2$) is much smaller than $\sigma_{e,Gain}$ ($\sim 10^{-16}$ cm$^2$). Therefore, at the lasing threshold, Eq. (3) becomes

$$\gamma = \frac{n_{1,Gain}}{n_{T,Gain}} = \gamma_0 + Xn_q. \qquad (4)$$

where $\gamma_0$ and X are:

$$\gamma_0 = \frac{\sigma_{a,Gain}}{\sigma_{e,Gain}}\left(1+\frac{L}{n_{T,Gain}\sigma_{a,Gain}}\right) \qquad (5)$$

$$X = \frac{\sigma_{a,q}}{n_{T,Gain}\sigma_{e,Gain}}. \qquad (6)$$

Eq. (4) states that for a microlaser in the absence of quenchers, we have $\gamma\left(n_q=0\right)=\gamma_0$, which corresponds to $\frac{n_{1,Gain}}{n_{T,Gain}}$ at the lasing threshold for an unquenched microlaser. $\gamma$ increases linearly with respective to the quencher density $n_q$, that is, in the presence of quenchers, the microlaser needs to receive a higher pumping intensity (or flux) and hence a higher $\frac{n_{1,Gain}}{n_{T,Gain}}$ to reach the lasing threshold.

Quenchers are produced by enzyme-catalyzed reactions. Here, we study the relation between the number of quenchers, $N_q$, and the number of analytes, $k$, inside a microlaser. First, we assume that the number of enzymes inside a microlaser, $D$, is always proportional to the number of analytes inside the microlaser ($D = C_1k$). In a simplified model, for a microlaser filled with liquid substrates, let $S$ be the number of the substrate molecules during reaction, $S_0$ be the initial number of the substrate molecules, $C_2$ be a reaction constant, and $T$ be the reaction time, we have

$$\frac{dS}{dt} = -C_2\frac{D}{V_0}S = -C_2C_1\frac{kS}{V_0}, \qquad (7)$$

$$n_q = \frac{N_q}{V_0} = \frac{S_0-S(T)}{V_0} = \frac{S_0}{V_0}\left(1-\frac{1}{e^{C_1C_2kT/V_0}}\right), \qquad (8)$$

where $V_0$ is the volume of the microlaser cavity. Eq. (7) states that in the enzyme-catalyzed reaction, where the substrate molecules are consumed, the reaction rate is proportional to enzyme number and substrate molecule number. Eq. (8) states that the number of quenchers is proportional to the product of the reaction, which can be expressed as the total consumed number of the substrate molecules.

From Eq. 8, we can get the form of $n_q$ for a fixed reaction time

$$n_q = A\left(1-\frac{1}{e^{Ek}}\right), \qquad (9)$$

where $A$ and $E$ are two constants. Plugging Eq. (9) into Eq. (4), we have

$$\gamma = \gamma_0 + XA\left(1-\frac{1}{e^{Ek}}\right), \qquad (10)$$

which connects the fraction of the gain molecules in the excited state at the lasing threshold (*i.e.*, $\gamma$) to the number of analytes captured by the microlaser through bio-interactions (such as antibody-antigen interactions).

Without losing generalizability, we use a four-energy-level laser system model. Therefore, at lasing threshold, we have

$$\gamma = \frac{\frac{I_{th}}{I_{sat,Gain}}}{\frac{I_{th}}{I_{sat,Gain}}+1} = \frac{\widetilde{I_{th}}}{\widetilde{I_{th}}+1}, \qquad (11)$$

where $I_{th}$ is the lasing threshold in a unit of flux (Jm$^{-2}$s$^{-1}$) for the laser. $I_{sat,Gain} = (\sigma_{a,Gain}\,\tau_{Gain})^{-1}$ is the saturation flux of the gain molecule, where $\sigma_{a,Gain}$ and $\tau_{Gain}$ are the absorption cross section and the lifetime of the gain molecule, respectively. $\widetilde{I_{th}}$ is the dimensionless normalized lasing threshold to simplify the expression. Let $I_{th}^{exp}$ be the experimentally measured pumping intensity (or flux), and $R_{eff}$ be an efficiency that considers the difference between the pumping intensity measured experimentally ($I_{th}^{exp}$) and the actual pumping intensity seen by the gain molecules, we have

$$\widetilde{I_{th}} = R_{eff}\frac{I_{th}^{exp}}{I_s} = \frac{I_{th}^{exp}}{I_{norm}}. \qquad (12)$$

The constant, $I_{norm}$, can be obtained using function fittings from experimental data.

Next, we connect the number of analytes on a microlaser with experimentally measured lasing threshold of a microlaser $I_{th}^{exp}$. First, notice a boundary condition that states that when the number of analytes on the microlaser is high, the microlaser breaks down ($I_{th} \to \infty$), we have

$$lim_{k\to\infty}(\gamma) = 1, \qquad (13)$$

which leads to a simplified form of Eq. (10), $i.e.$,

$$\gamma = \gamma_0 + (1-\gamma_0)\left(1 - \frac{1}{e^{Ek}}\right). \qquad (14)$$

We use $I_0$ to represent the experimental lasing threshold of a microlaser in the absence of quenchers. From Eqs. (11), (12) and (14), we have

$$\gamma_0 = \frac{\frac{I_0}{I_{norm}}}{1+\frac{I_0}{I_{norm}}}. \qquad (15)$$

Using Eqs. (11), (12), (14), and (15), the expression of experimentally measured lasing threshold of a microlaser with respect to the number of analytes inside the microlaser is

$$I_{th}^{exp} = (I_{norm} + I_0)e^{Ek} - I_{norm}, \qquad (16)$$

which in turn leads to

$$k = \frac{ln\left(\frac{I_{th}^{exp}+I_{norm}}{I_0+I_{norm}}\right)}{E}. \qquad (17)$$

Eq. (17) shows how to obtain the number of analytes associated with a microlaser using the experimentally measured lasing threshold. $I_{norm}$ and $E$ can be obtained using function fittings from experimental data. So far, $k$ is directly connected to pumping intensity (flux), which has a unit of Jm$^{-2}$s$^{-1}$. In our experiment, we use a pulsed nanosecond laser as the excitation source. Therefore, it is easier to describe the pumping intensity per pulse in terms of pumping energy density per pulse. Pumping intensity is pumping energy density divided by $\delta t_{pulse}$, the pulse width that carries a unit of time. In the main text, we use $I_{th}^{exp}$, $I_{norm}$, and $I_0$ to refer to the pumping energy density per pulse, but they should be understood as the pumping intensity per pulse. Eq. (17) is still valid, since the term, $\delta t_{pulse}$, is cancelled out in Eq. (17).

**S1.2. From a single microlaser to a microlaser ensemble**

Previously, we have described the relation between the lasing threshold of a single microlaser and the number of analytes captured by the microlaser. Now we move to the microlaser ensemble where the number of analytes on each microlaser is different due to the random binding processes during analyte capturing. Considering that we have a large number of the microlasers, from the central limit theorem, we assume that the distribution of the analytes on the microlasers follows the Gaussian distribution. For microlaser $j$ in the ensemble, $k_j$ is the number of analytes on this microlaser, we have:

$$k_j \sim \text{Gaussian}(\mu_c, \sigma_c), \quad (18)$$

where $\mu_c$ and $\sigma_c$ are the mean and the standard deviation of the molecule number distribution, and they are dependent on $C$, the concentration of the analytes in the liquid sample. From Eq. (17), with the assumption described in Eq. (18), we present a method of measuring analyte number distribution on a microlaser ensemble by measuring its lasing threshold distribution as described as follows.

Under a fixed pumping intensity (flux), we can count the number of the microlasers that lase to calculate the lasing fraction in the microlaser ensemble. These lasing microlasers all have a lasing threshold smaller than or equal to the pumping intensity (flux), as illustrated in Fig. S1(a). Therefore, by changing the pumping intensity (flux), we probe the lasing threshold distribution in the microlaser ensemble. According to Eq. (18), probing the lasing microlaser distribution is equivalent to probing the distribution of the analytes in the microlaser ensemble, which is illustrated in Fig. S1(b), that is, the microlasers that lase have the number of analytes smaller than or equal to a specific value. Mathematically, it can be stated as the lasing fraction and pumping intensity (flux) follow the following equation:

$$Lf(i) = \phi_{cumu}(k_i; \mu_c, \sigma_c), \quad (19)$$

where

$$k_i = \frac{ln\left(\frac{I_{pump}^{exp}(i)+I_{norm}}{I_0+I_{norm}}\right)}{E} \quad (20)$$

and $\phi_{cumu}$ stands for the standard Gaussian cumulative distribution function, and $i$ denotes a specific pumping intensity (flux). By using multiple pumping intensities, we can obtain a set of data, [($Lf(i)$, $I_{pump}^{exp}(i)$)], which can be used to fit the standard Gaussian cumulative distribution and obtain $\mu_c$ and $\sigma_c$. Finally, by using the one-to-one correspondence between the analyte distribution ($\mu_c$ and $\sigma_c$) and the analyte concentration $C$ in the liquid solution, we can obtain the analyte concentration. This correspondence between ($\mu_c$ and $\sigma_c$) of a microlaser ensemble and the analyte concentration can be established by a calibration curve.

All symbols used in this article are listed in Table S1.

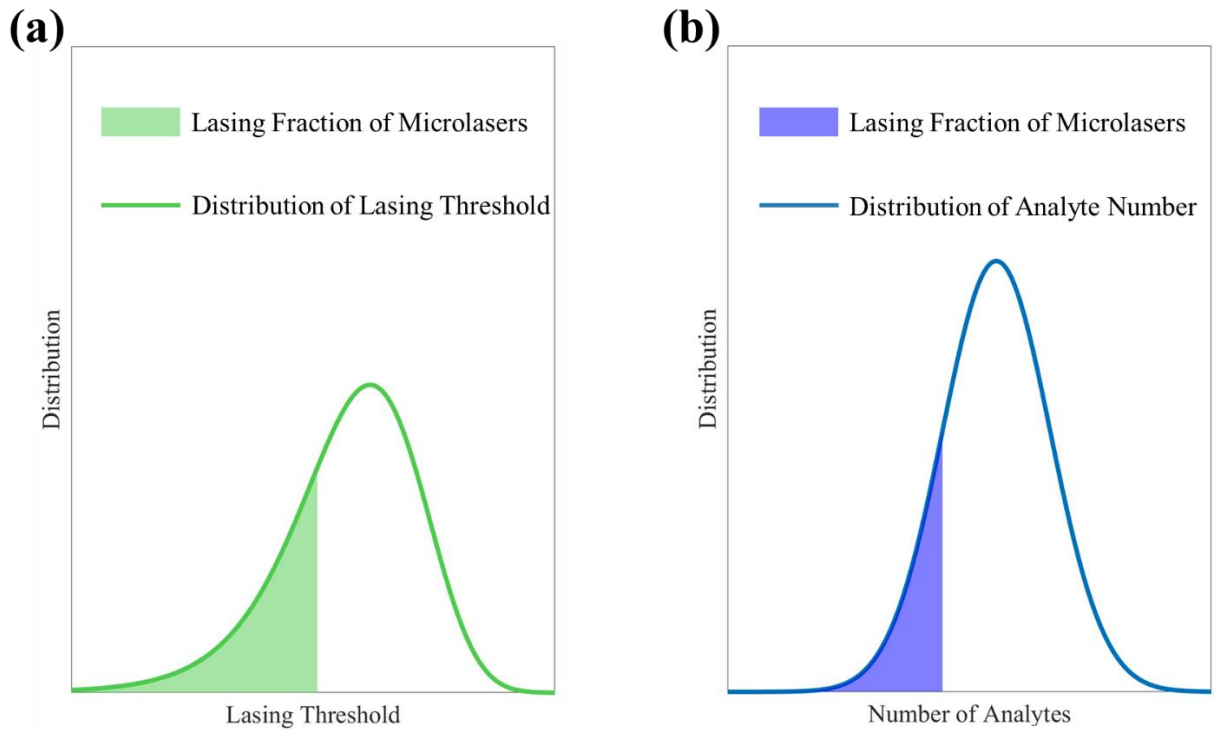

Figure S1. (a) The curve shows the lasing threshold distribution in a microlaser ensemble. The filled area represents the lasing fraction of the microlasers in the ensemble when pumped at a specific pumping intensity. (b) The curve shows analyte number distribution of the same ensemble. The blue-shaded area represents the same bright microlasers (lasing fraction) when pumped at a specific pumping intensity. Note that pumping at one pumping intensity is equivalent to probing analyte number smaller than a specific value.

<div align="center">**S2. Sensitivity Analysis**</div>

In the following discussion, we compare the sensitivity between the microlaser quenching method and conventional fluorescence-based method.

## S2.1. Sensitivity in fluorescence-based detection

The conventional fluorescence-based biosensing method is to attach a fluorophore (such as dyes, quantum dots, and engineered fluorescent molecules, *etc.*) to the analyte, which is subsequently captured to a biosensor surface or body. The fluorescence signal is used to quantify the analyte. For simplicity, we assume that each analyte has only one fluorophore attached to it. When $k$ analytes are attached to the biosensor, the signal intensity, represented by the total photon emission rate $P_{Fl}$, is

$$P_{Fl} = \frac{Q_e}{\tau_{Fl}} N_{1,Fl}, \qquad (21)$$

where $N_{1,Fl}$ is the number of fluorophores in the excited state. $\tau_{FL}$ and $Q_e$ are the lifetime and quantum yield of the fluorophore, respectively. $N_{1,Fl}$ is related to the total number of fluorophore $N_{T,\,Fl}$ ($=k$) by

$$I_{p,Fl}\sigma_{a,Fl}\left(N_{T,Fl} - N_{1,Fl}\right) = I_{p,Fl}\sigma_{a,Fl}\left(k - N_{1,Fl}\right) = \frac{N_{1,Fl}}{\tau_{Fl}}, \qquad (22)$$

where $I_{p,Fl}$ is the flux of the excitation light and $\sigma_{a,Fl}$ is the absorption cross section of the fluorophore. Combining Eqs. (21) and (22), we have

$$P_{Fl} = \frac{Q_e}{\tau_{Fl}} \frac{I_{p,Fl}\sigma_{a,Fl}\tau_{Fl}}{1 + I_{p,Fl}\sigma_{a,Fl}\tau_{Fl}} k = \frac{Q_e}{\tau_{Fl}} \frac{\frac{I_{p,FL}}{I_{sat,Fl}}}{1 + \frac{I_{p,FL}}{I_{sat,Fl}}} k, \qquad (23)$$

where $I_{sat,Fl} = (\sigma_{a,Fl}\tau_{Fl})^{-1}$ is the saturation flux of the fluorophore. Practically, only a portion of the emitted photons can be collected. Therefore,

$$P_{Fl}^{Exp} = R_{Collect} \frac{Q_e}{\tau_{Fl}} \frac{\frac{I_{p,Fl}}{I_{sat,Fl}}}{1 + \frac{I_{p,Fl}}{I_{sat,Fl}}} k, \qquad (24)$$

where $P_{Fl}^{Exp}$ is experimentally measured photon emission rate and $R_{Collect}$ is the photon collection efficiency.

The sensitivity of fluorescence-based detection, $S_{Fl}$, is defined as the emitted photon rate change in response to the change of the analyte number, *i.e.*,

$$S_{Fl} = \frac{dP_{Fl}^{Exp}}{dk} = R_{Collect} \frac{Q_e}{\tau_{Fl}} \frac{\frac{I_{p,Fl}}{I_{sat,Fl}}}{1 + \frac{I_{p,Fl}}{I_{sat,Fl}}}, \qquad (25)$$

which gives the photon emission rate change when the number of analyte changes from $k$ to $k+1$.

## S2.2. Sensitivity in microlaser quenching-based detection

For simplicity and for a fair comparison with the fluorescence-based detection, here we first assume that each analyte has only one quencher molecule attached to it, which is similar to the fluorescence-based detection discussed previously, where each analyte has only one fluorophore attached to it (see later in Section 2.5 for the case where multiple quenchers are generated through enzyme-catalyzed reactions). According to Eq. (11), the lasing threshold of the microlaser, $I_{th}$, can be expressed as

$$I_{th} = I_{sat,Gain} \frac{\gamma}{1-\gamma}. \qquad (26)$$

where $I_{sat,Gain}$ is the saturation photon flux of the gain molecule and $\gamma$ is given by Eq. (4)

$$\gamma = \gamma_0 + X n_q = \gamma_0 + X \frac{k}{V_0}, \qquad (27)$$

where $V_0$ is the microlaser volume that converts the number of quenchers (hence analytes) into the density of quenchers (analytes).

The sensitivity of laser quencher-based detection is defined as the lasing threshold change in response to the change of the analyte number from $k$ to $k+1$, i.e.,

$$\frac{dI_{th}}{dk} = I_{sat,Gain} \left( \frac{1}{(1-\gamma)^2} \right) \frac{X}{V_0}. \qquad (28)$$

To further simplify Eq. (28), without losing generality we assume that $\sigma_{a,q} = \sigma_{e,Gain}$. As a result, Eq. (28) becomes

$$\frac{dI_{th}}{dk} = \frac{I_{sat,Gain}}{V_0 n_T} \frac{1}{\left( 1 - \gamma_0 - \frac{k}{V_0 n_{T,Gain}} \right)^2}. \qquad (29)$$

Note that the sensitivity defined in Eq. (29) has a unit of $Jm^{-2}s^{-1}$ (excitation photon flux), which is different from $S_{Fl}$, which carries a unit of $s^{-1}$. To make a direct comparison, we use the following equations to convert the photon flux into the photon absorption rate (in a unit of $s^{-1}$).

$$\Sigma_{a,Gain}^{Pump} = \sigma_{a,Gain}^{Pump} V_0 n_{T,Gain}, \qquad (30)$$

$$P_{th} = I_{th} \Sigma_{a,gain}^{Pump}. \qquad (31)$$

$\Sigma_{a,Gain}$ is the total absorption cross section of the gain molecules inside the microlaser at the pump wavelength. $P_{th}$ is the interaction rate between the pump photons and the entire gain molecules within the microlaser (or pump photon absorption rate at the lasing threshold). According to the above conversion, the sensitivity of the microlaser quenching-based method, $S_{Laser}$, can be defined

$$S_{Laser} = \frac{dP_{th}}{dk} = \frac{1}{\tau_{Gain}} \frac{1}{\left( 1 - \gamma_0 - \frac{k}{V_0 n_{T,Gain}} \right)^2}. \qquad (32)$$

From Eq. (32), we can see that when $k$ (the number of analytes and hence the number of quenchers) is small as compared to the total number of gain molecules, which is on the order of $10^{10}$ for a microlaser, $S_{Laser}$ remains a constant. However, when the ratio between the number of the analytes and the total number of the gain molecules approaches $\gamma_0$, $S_{Laser}$ starts to increase significantly with respect to $k$.

## S2.3. Sensitivity comparison

Based on Eqs. (25) and (32), the sensitivity ratio, $R_{Sens}$ between the microlaser quenching- and fluorescence-based methods can be expressed as

$$R_{Sens} = \frac{S_{Laser}}{S_{Fl}} = \frac{1}{R_{Collect}} \frac{1}{Q_e} \frac{\tau_{Fl}}{\tau_{Gain}} \frac{1 + \frac{I_{p,Fl}}{I_{sat,Fl}}}{\frac{I_{p,Fl}}{I_{sat,Fl}}} \frac{1}{\left(1 - \gamma_0 - \frac{k}{V_0 n_{T,Gain}}\right)^2}. \qquad (33)$$

For a numerical comparison, we use the values in Table S2. Accordingly, Eq. (13) becomes

$$R_{Sens} = \frac{1}{1.25 \times 10^{-6}} \frac{1}{\left(1 - 0.05 - \frac{k}{6 \times 10^{10}}\right)^2}. \quad (34)$$

When k is $<<10^{10}$, $R_{Sens}$ is $\sim 10^6$.

Note that in all the above discussions, we have assumed that the absorption of the excitation light, $I_{p,Fl}$, in the fluorescence-based method is linear with respect to the analyte concentration (or fluorophore concentration). That is, the optical absorption of the excitation light follows the Beer-Lambert law. However, the reality is that when $k$ is large (so is the analyte/fluorophore concentration), the Beer-Lambert law breaks down. As a result, the increase in the fluorescence signal becomes sub-linear with respect to the increase in the analyte. Mathematically, Eq. (23) should be changed to

$$P_{Fl} = \frac{Q_e}{\tau_{Fl}} \frac{\frac{I_{p,FL}}{I_{sat,Fl}}}{1 + \frac{I_{p,FL}}{I_{sat,Fl}}} \frac{k}{1 + \alpha k} \quad , \qquad (35)$$

where $\alpha$ is a parameter that accounts for the non-linearity of optical absorption at a high $k$. At a low $k$, Eq. (35) is reduced to Eq. (23). When $k$ is very large, $P_{Fl}$ increases sub-linearly. The sensitivity described in Eq. (5) becomes

$$S_{Fl} = \frac{Q_e}{\tau_{Fl}} \frac{\frac{I_{p,FL}}{I_{sat,Fl}}}{1 + \frac{I_{p,FL}}{I_{sat,Fl}}} \frac{1}{(1 + \alpha k)^2}. \qquad (36)$$

Thus, the sensitivity, $S_{Fl}$ in Eq. (25), is no longer constant and it decreases with respect to $k$.

Similarly, for the microlaser quenching method, when the number of analytes (and hence the number of quencher) is high, Eq. (27) should become

$$\gamma = \gamma_0 + X \frac{1}{V_0} \frac{k}{1 + \beta k}, \qquad (37)$$

where $\beta$ is a parameter that accounts for the non-linearity of optical absorption of the quenchers at a large $k$ (or a large number of quenchers). Accordingly, the sensitivity described in Eq. (32)

$$S_{Laser} = \frac{1}{\tau_{Gain}} \frac{1}{\left(1 - \gamma_0 - \frac{k}{V_0 n_{T,Gain}}\right)^2} \frac{1}{(1 + \beta k)^2} \propto \frac{1}{(1 - \varepsilon k)^2} \frac{1}{(1 + \beta k)^2}, \qquad (38)$$

where

$$\varepsilon = \frac{1}{(1 - \gamma_0) V_0 n_{T,Gain}}. \qquad (39)$$

Thus, the first term in Eq. (38) counteracts the second term to mitigate the sensitivity decreasing effect at a large $k$.

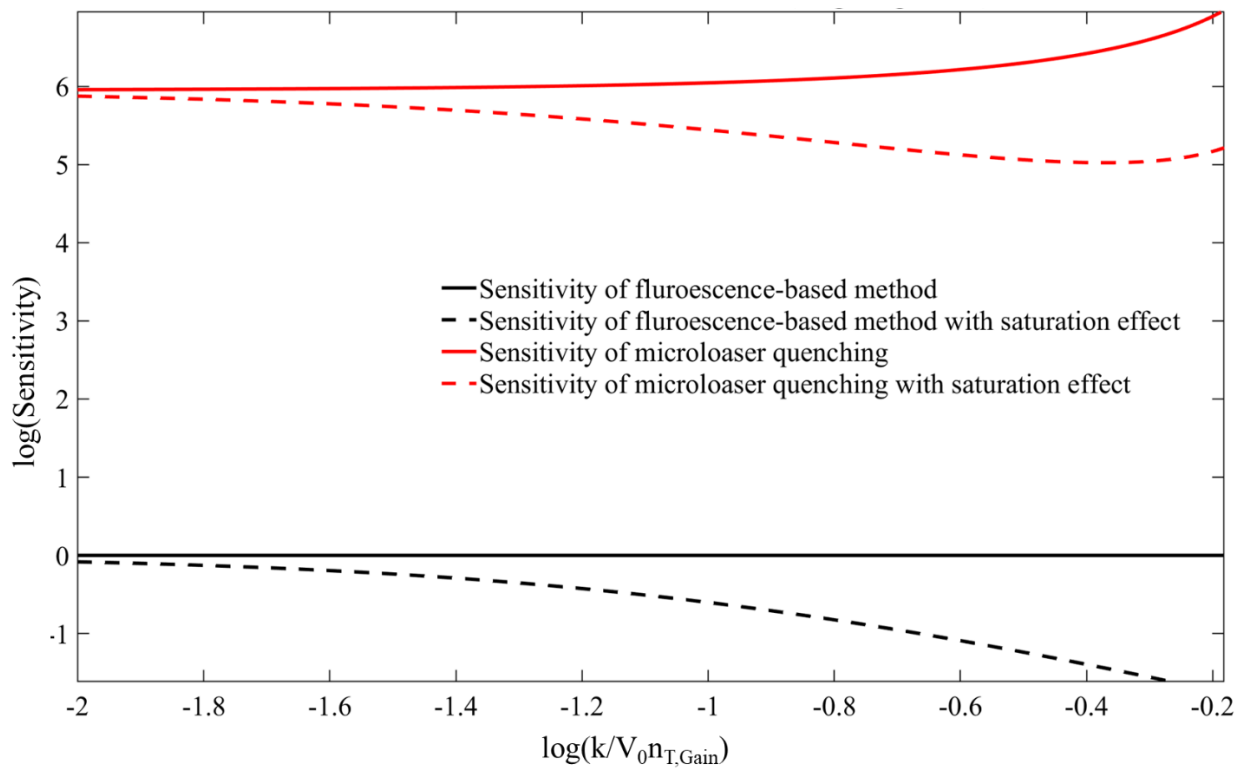

Figure S2. An illustration of sensitivity comparison between the microlaser quenching method with the fluoescence-based method, where the sensitivity of the fluorescence-based method is set to be 1. The solid line shows the sensitivity of each method without considering the saturation effect. The sensitivity of the microlaser quenching method is calculated using Eq. (34). The dashed lines show the sensitivity of each method with saturation effect. For the fluorescence based method, the sensitivity is calculated using Eq. (36), whereas the sensitivity for microlaser quenching method is shown in Eq. (38). The saturation factor for the two methods, $\alpha$ and $\beta$, as shown in Eqs. (36) and (38), are set both to be $10V_0\,n_{T,Gain}$.

To further observe the superior sensitivity of the microlaser quenching method over the fluorescence-based method, and the mitigation effect of microlaser quenching against the saturation effect (absorption), we plot the sensitivity comparison of the two methods with the numerical values we used for Eq. (34). We set the sensitivity value for the fluorescence-based method without saturation effects to be 1, and the sensitivity for the microlaser quenching method without saturation effect is then in the form of Eq. (34). For the comparison of sensitivity with saturation effect, the sensitivity for fluorescence-based value is shown in Eq. (36), where we set the saturation factor $\alpha$ to be $10V_0\,n_{T,Gain}$. For a fair comparison, we set $\beta$, the saturation effect for the microlaser quenching method in Eq. (38), to be the same as $\alpha$. Using Eqs. (36) and (38), we can calculate the sensitivity of the two methods with the saturation effect. Note that for the microlaser quenching method, the mitigation effect averts the sensitivity from further dropping at high analyte numbers (where the derivative of the sensitivity curve becomes positive).

## S2.4. Distribution of sensing signal from an ensemble of sensors

When an ensemble of fluorescence-based sensors and an ensemble of microlaser-based sensors are incubated with the same analyte of a fixed concentration, both ensembles have the same analyte distribution over their respective sensors within the ensemble. Using $N(\mu_c, \sigma_c)$ to represent the Gaussian distribution of analytes in both ensembles, the fluorescence signal, $P_{Fl}$, has a Gaussian distribution denoted by $N(\mu_c^{Fl}, \sigma_c^{Fl})$, where:

$$\mu_c^{Fl} = R_{collect} \frac{Q_e}{\tau_{Fl}} \frac{I_p \sigma_{a,Fl} \tau_{Fl}}{1 + I_p \sigma_{a,Fl} \tau_{Fl}} \mu_c, \text{ (40)}$$

$$\sigma_c^{Fl} = R_{collect} \frac{Q_e}{\tau_{Fl}} \frac{I_p \sigma_{a,Fl} \tau_{Fl}}{1 + I_p \sigma_{a,Fl} \tau_{Fl}} \sigma_c. \text{ (41)}$$

Similarly, the microlaser based sensing signal, $P_{th}$, has a Gaussian distribution denoted by $N(\mu_c^{th}, \sigma_c^{th})$, where:

$$\mu_c^{th} = \frac{1}{\tau_{Gain}} \frac{1}{(1-\gamma_0)^2} \mu_c, \quad \text{(42)}$$

$$\sigma_c^{th} = \frac{1}{\tau_{Gain}} \frac{1}{(1-\gamma_0)^2} \sigma_c. \quad \text{(43)}$$

Note that for simplicity, we assume that $k \ll V_0 n_{T,Gain}$, which is usually true in most cases. Eqs. (40) – (43) show how the distribution of the analytes over the sensors is transformed to the distribution of the sensing signal. As discussed previously in S2.3, $\frac{1}{\tau_{Gain}} \frac{1}{(1-\gamma_0)^2}$ is much larger than $R_{collect} \frac{Q_e}{\tau_{Fl}} \frac{I_p \sigma_{a,Fl} \tau_{Fl}}{1 + I_p \sigma_{a,Fl} \tau_{Fl}}$. Therefore, the mean value and the spread of the signal distribution in the microlaser-quenching method is much larger than the counterparts in the fluorescence-based method, thus making the microlaser-quenching method much more sensitive to detect the analyte distribution within the sensor ensemble.

**S2.5. Sensitivity in microlaser quenching-based detection with enzyme amplification**

In the direct labelling method discussed in the previous sections, we set the number of quenchers to be equal to the number of analytes (*i.e.*, $N_q=k$). Therefore, we have

$$\frac{dP_{th}}{dk} = \frac{dP_{th}}{dN_q}. \qquad (44)$$

For the enzyme-amplification method, using $P_{th}^{enzyme}$ to denote the sensing signal, we have:

$$S_{Laser}^{Enzyme} = \frac{dP_{th}^{Enzyme}}{dk} = \frac{dP_{th}^{Enzyme}}{dN_q}\frac{dN_q}{dk}. \qquad (45)$$

We further use $C_{amp}$ to represent the number of quenchers produced by one analyte, *i.e.*,

$$C_{amp} = \frac{dN_q}{dk}. \qquad (46)$$

Therefore, the sensitivity of the microlaser quenching-based method is increased by a factor of $C_{amp}$ when enzyme-catalyzed reaction is employed, *i.e.*,

$$S_{Laser}^{Enzyme} = C_{amp}S_{Laser}. \qquad (47)$$

**S3. Raw lasing fraction data prior to normalization**

Here, we show how the normalization removes the "bad" microlasers. Assume that we have two sets of MEs. One set is incubated in a negative control sample, whereas the other one is incubated in a positive group with a certain analyte concentration (say, 1 ng/mL). Then for the two sets of MEs, the raw, unnormalized lasing fractions are:

$$Lf(Neg)_{raw} = \frac{N_{bright}(Neg)}{N_{good} + N_{bad}}, \qquad (48)$$

$$Lf(Pos)_{raw} = \frac{N_{bright}(Pos)}{N_{good} + N_{bad}}, \qquad (49)$$

where $Lf(Neg)_{raw}$ is the lasing fraction of the ME in the negative control sample (0 pg/mL) and $N_{bright}(Neg)$ is the number of bright microlasers in the ME under a given pumping energy density. $Lf(Pos)_{raw}$ is the lasing fraction of the ME in a positive sample and $N_{bright}(Pos)$ is the number of bright microlasers in the ME under a given pumping energy density. $N_{good}$ and $N_{bad}$ refer to the number of "good" microlasers (those with lasing thresholds smaller than 400 µJ/mm$^2$) and "bad" microlasers (those with lasing thresholds larger than 400 µJ/mm$^2$), respectively. The total number of microlasers (*i.e.*, $N_{good} + N_{bad}$) for each ME is kept very similar in our work, *i.e.*, ~1000.

In the normalization step, we have the normalized lasing fraction under the same pumping energy density, $Lf(Pos)_{norm}$, expressed as:

$$Lf(Pos)_{norm} = \frac{Lf(Pos)_{raw}}{Lf(Neg)_{raw}} = \frac{N_{bright}(Pos)}{N_{bright}(Neg)}. \qquad (50)$$

As a result, the dependency on the number of good or bad microbeads is removed.

**S3. Example of the calibration curves in fluorescence-based immunoassay**

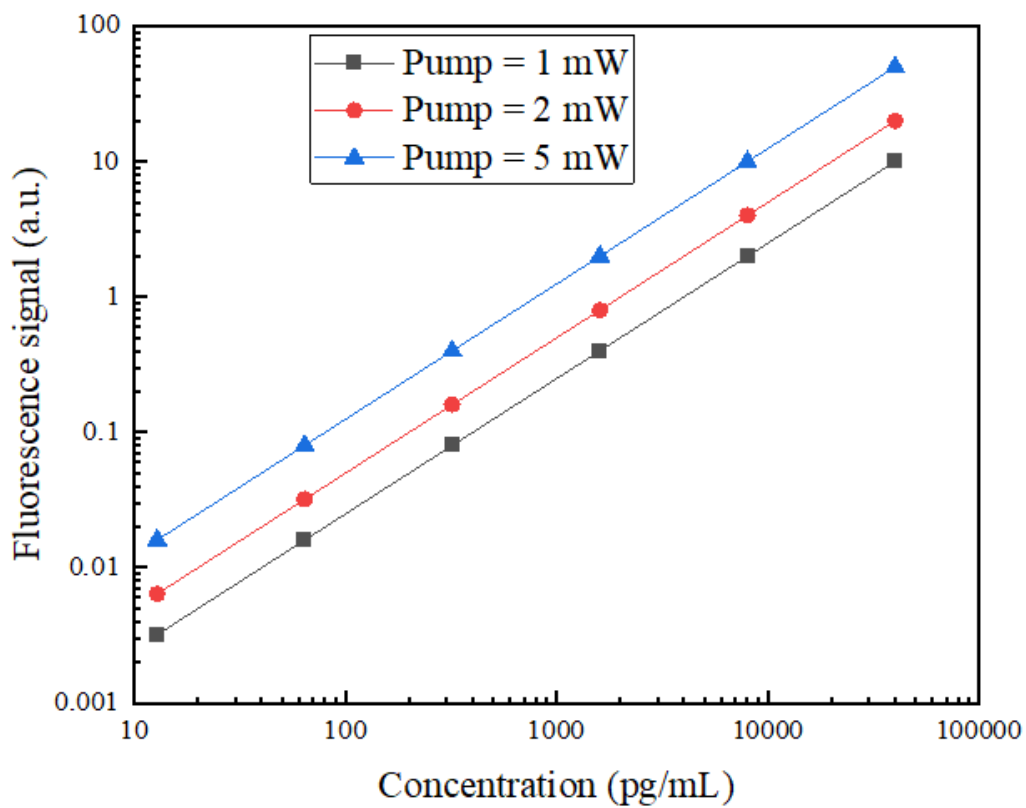

Figure S3. Example of the three calibration curves obtained for the same fluorescence-based immunoassay under three different pumping powers. These three calibration curves are essentially identical, since they can be reduced to one curve with a rescaling factor. For example, the blue and red curves become the black curve when their fluorescence signals are divided by a factor of 5 and 2, respectively.

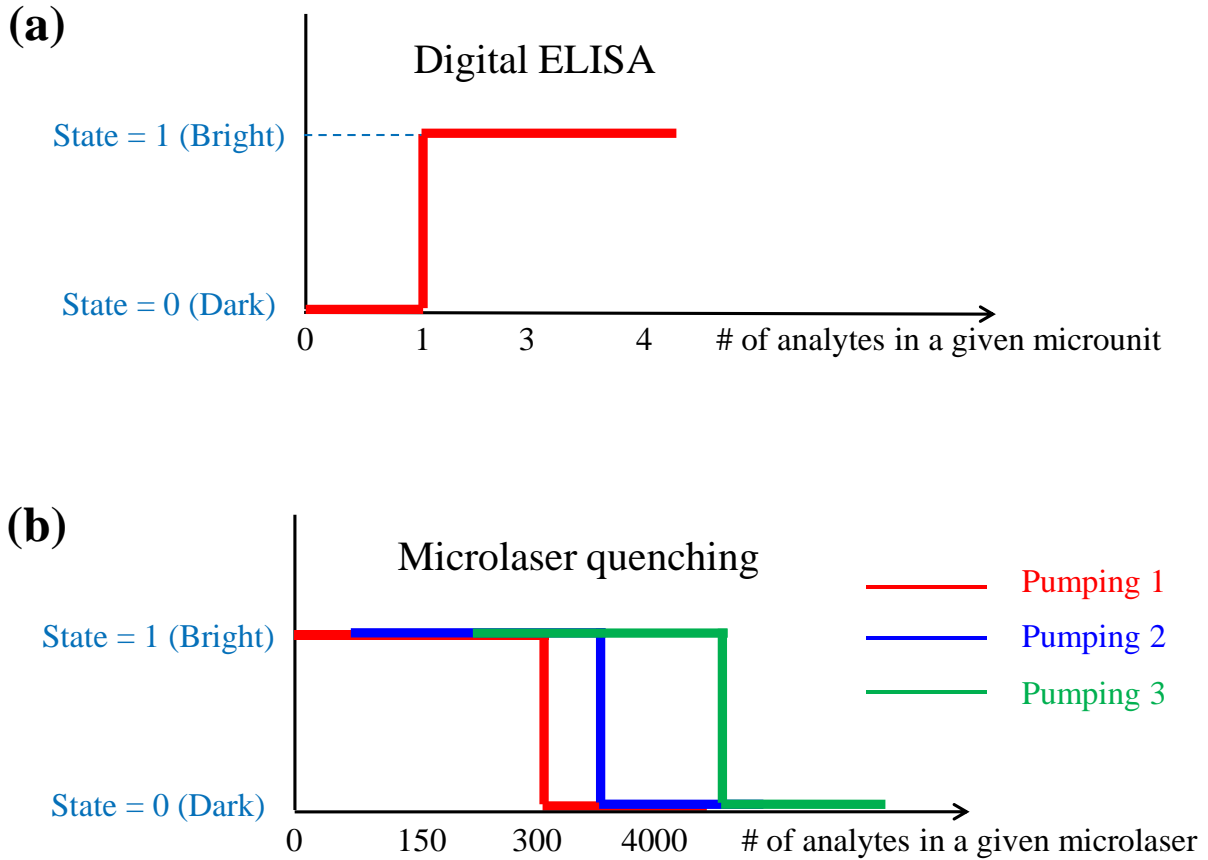

Figure S4. Comparison between digital ELISA (a) and our method (b) regarding how a microunit is categorized. (a) In digital ELISA, when there is zero analyte in a microunit, this microunit is categorized as "Dark". When there is one analyte in a microunit, this microunit is categorized as "Bright". When there are two or more analytes in a microunit, this microunit is still categorized as "Bright". Therefore, for each microunit, digital ELISA can tell one analyte from zero analyte in a microunit, but it saturates when two or more analytes are in a microunit. The cut-off value for "Dark" and "Bright" is one. (b) In our method, a microunit (*i.e.*, microlaser) is categorized as "Bright" when the number of analytes is below a certain cut-off value under a given pumping energy density. However, the cut-off value can be tuned by changing the pumping energy densities. Therefore, the dynamic range can be extended. Note that in the x-axis, the number of analytes increases from left to right, but is marked arbitrarily for illustration purposes and is not to scale.

| Table S1. List of symbols. | | | |
|---|---|---|---|
| **Symbol** | **Definition** | **Symbol** | **Definition** |
| $n_{1,Gain}$ | the density of the laser gain molecules in the excited state | $V_0$ | the volume of the microlaser cavity |
| $n_{T,Gain}$ | the total density of the laser gain molecules in the microlaser | $I_{sat,Gain}$ | the saturation photon flux of the gain molecule |
| $n_q$ | the quencher density at the lasing wavelength | $I_{sat,Fl}$ | the saturation flux of the fluorophore |
| $\sigma_{e,Gain}$ | the emission cross section of the laser gain molecules at the lasing wavelength | $C$ | the concentration of the analytes in the liquid sample |
| $\sigma_{a,Gain}$ | the absorption cross section of the laser gain molecules at the lasing wavelength | $k_i$ | the number of analytes on the $i_{th}$ microlaser |
| $\sigma_{a,q}$ | the quencher's absorption cross section at the lasing wavelength | $\mu_c$ | the mean of the analyte molecule number distribution |
| $L$ | the intrinsic per single-trip laser cavity loss in the absence of quenchers | $\sigma_c$ | the standard deviation of the analyte molecule number distribution |
| $k$ | the number of analytes inside a microlaser. | $\mu_c^{Fl}$ | the mean of the signal intensity distribution for the fluorescence-based method |
| $\gamma_0$ | $\frac{n_{1,Gain}}{n_{T,Gain}}$ at the lasing threshold for an unquenched microlaser | $\sigma_c^{Fl}$ | the standard deviation of the signal intensity distribution for the fluorescence-based method |
| $\gamma$ | $\frac{n_{1,Gain}}{n_{T,Gain}}$ at the lasing threshold for a quenched microlaser | $\mu_c^{th}$ | the mean of the signal intensity distribution for the microlaser quenching method |
| $\tau_{Gain}$ | the lifetime of the gain molecules | $\sigma_c^{th}$ | the standard deviation of the signal intensity distribution for the microlaser quenching method |
| $\tau_{Fl}$ | the lifetime of the fluorophores | $P_{Fl}$ | the total photon emission rate in the fluorescence-based method |
| $N_q$ | the number of quenchers inside a microlaser | $P_{th}^{exp}$ | the experimentally measured photon emission rate in the fluorescence-based method |
| $S$ | the number of the substrate molecules during reaction | $P_{th}$ | the pump photon absorption rate at the lasing threshold |
| $S_0$ | the initial number of the substrate molecule | $N_{1,Fl}$ | the number of fluorophores in the excited state |
| $C_2$ | the reaction constant for enzyme-substrate reaction | $P_{th}^{enzyme}$ | the sensing signal for the enzyme amplification method |
| $T$ | the reaction time for enzyme-substrate reaction | $Q_e$ | the quantum yield of the fluorophore |
| $I_t$ | the lasing threshold in units of flux for the microlaser | $I_{p,Fl}$ | the flux of the excitation light in the fluorescence-based method |
| $I_{th}^{exp}$ | the experimentally measured pumping intensity (or flux) | $\sigma_{a,Fl}$ | the absorption cross section of the fluorophore |
| $R_{Collect}$ | the photon collection efficiency of the photodetector in experiments | $S_{Fl}$ | the emitted photon rate change in response to the change of the analyte number |
| $R_{Sens}$ | the sensitivity ratio between the microlaser quenching and fluorescence-based methods | $\Sigma_{a,Gain}$ | the total absorption cross section of the gain molecules inside the microlaser at the pump wavelength |

| Table S2. Parameters and their values used in Eq. (34). | | | |
|---|---|---|---|
| **Parameter** | **Value** | **Parameter** | **Value** |
| $R_{Collect}$ | 0.5 | $Q_e$ | 1 |
| $\tau_{Fl}$ | $10^{-8}$ s | $\sigma_{a,Fl}$ | $10^{-16}$ cm$^2$ |
| $I_p$ | 1 W cm$^{-2}$ = $2.77 \times 10^{21}$ m$^{-2}$s$^{-1}$ (Note 1) | $\tau_{Gain}$ | $10^{-8}$ s |
| $\gamma_0$ | 0.05 (Note 2) | $V_0$ | $10^{-15}$ m$^3$ (Note 3) |
| $n_{T,Gain}$ | 0.1 mol L$^{-1}$ = $6 \times 10^{25}$ m$^{-3}$ (Note 4) | $\beta$ | $6 \times 10^{11}$ |
| $\alpha$ | $6 \times 10^{11}$ | | |
| Note 1: At wavelength of 550 nm. | | | |
| Note 2: Typical value for a dye-based microlaser. | | | |
| Note 3: A 10 μm x 10 μm x 10 μm cubic microlaser. | | | |
| Note 4: Typical dye doping concentration (or density) in a microlaser. | | | |
| $\alpha$ | the absorption parameter for the non-linearity of optical absorption for the fluorescence-based biosensor | $\beta$ | the absorption parameter for the non-linearity of optical absorption for the microlaser quenching method |

**References**

1.  M. Aas, Q. Chen, A. Jonáš, A. Kiraz, and X. Fan, "Optofluidic FRET lasers and their applications in novel photonic devices and biochemical sensing," IEEE J. Sel. Top. Quantum Electron. **22**, 188-202 (2015).
2.  S. Lacey, I. M. White, Y. Sun, S. I. Shopova, J. M. Cupps, P. Zhang, and X. Fan, "Versatile opto-fluidic ring resonator lasers with ultra-low threshold," Opt. Express **15**, 15523-15530 (2007).
3.  H.-J. Moon, Y.-T. Chough, and K. An, "Cylindrical microcavity laser based on the evanescent-wave-coupled gain," Phys. Rev. Lett. **85**, 3161 (2000).
4.  A. E. Siegman, *Lasers* (University science books, 1986).